\documentclass[twocolumn,floatfix,linenubers]{aastex63}

\usepackage{amsmath,amstext}
\usepackage[T1]{fontenc}
\usepackage[utf8]{inputenc}
\usepackage[figure,figure*]{hypcap}
\usepackage{amsmath,amssymb,amsfonts,graphicx}
\usepackage{color}
\usepackage{graphicx}
\usepackage{tabularx}
\usepackage{footnote}
\usepackage{subfigure}

\newcommand{\teff}{$T_\textrm{eff}$}
\newcommand{\teffp}{$T_{1}^\textrm{eff}$}
\newcommand{\teffs}{$T_{2}^\textrm{eff}$}
\newcommand{\logg}{$\log g$}
\newcommand{\feh}{[Fe/H]}
\newcommand{\porb}{$P_\textrm{orb}$}
\newcommand{\chitwo}{$\chi^{2}$}
\newcommand{\halpha}{H$\alpha$}
\received{}
\revised{}
\accepted{}
\submitjournal{ApJ}

\begin{document} 

\title{Long-term photometric and low-resolution spectroscopic analysis of five contact binaries}

\correspondingauthor{Panchal, A.}
\email{alaxender@aries.res.in}

\author{Panchal, A.}
\affiliation{Aryabhatta Research Institute of observational sciencES (ARIES). Nainital, Uttrakhand, India.}
\affiliation{Department of Physics, DDU Gorakhpur University, Gorakhpur, India.}
%\email{alaxender@aries.res.in}

\author{Joshi, Y. C.}
\affiliation{Aryabhatta Research Institute of observational sciencES (ARIES). Nainital, Uttrakhand, India.}

\author{De Cat, Peter}
\affiliation{Royal Observatory of Belgium, Ringlaan 3, B-1180 Brussel, Belgium.}

\author{Tiwari, S. N.}
\affiliation{Department of Physics, DDU Gorakhpur University, Gorakhpur, India.}

%\collaboration{1}{(AAS Journals Data Scientists collaboration)}

%\author{Butler Burton}
%\affiliation{Leiden University}
%\affiliation{AAS Journals Associate Editor-in-Chief}
%\nocollaboration{1}

%\altaffiliation{AASTeX v6+ programmer}

%\collaboration{1}{(LaTeX collaboration)}

%\affiliation{AAS Director of Publishing}
%\affiliation{American Astronomical Society \\
%1667 K Street NW, Suite 800 \\

%\author{Scott Chernoff}
%\affiliation{IOP Publishing, Washington, DC 20005}




%\linenumbers

\begin{abstract}

{ A photometric and spectroscopic investigation is performed on five W Ursae Majoris eclipsing binaries (EWs) J015818.6+260247 (hereinafter as J0158b), J073248.4+405538 (hereinafter as J0732), J101330.8+494846 (hereinafter as J1013), J132439.8+130747 (hereinafter as J1324) and J152450.7+245943 (hereinafter as J1524). The photometric data are collected with the help of the 1.3\,m Devasthal Fast Optical Telescope (DFOT), the 1.04\,m Sampurnanand Telescope (ST) and the TESS space mission. The low-resolution spectra of the 4\,m Large Sky Area Multi-Object Fiber Spectroscopic Telescope (LAMOST) are used for spectroscopic analysis. The orbital period change of these systems is determined using our and previously available photometric data from different surveys. The orbital period of J1013 and J1524 is changing with a rate of $-2.552(\pm0.249)\times 10^{-7}$ days $yr^{-1}$ and $-6.792(\pm0.952)\times 10^{-8}$ days $yr^{-1}$, respectively, while others do not show any orbital period change. The orbital period change of J1013 and J1524 corresponds to a mass transfer rate of $2.199\times10^{-7} M_{\odot}\,yr^{-1}$ and $6.151\times10^{-8}M_{\odot}\,yr^{-1}$ from the primary to the secondary component in these systems. It is likely that angular momentum loss via magnetic braking may also be responsible for the observed orbital period change in the case of J1524. All systems have a mass-ratio lower than 0.5, except J0158b with a mass-ratio of 0.71. All the systems are shallow type contact binaries. The J0158b and J1524 are A-subtype while others are W-subtype. The $H_{\alpha}$ emission line region is compared with template spectra prepared using two inactive stars with the help of STARMOD program. The J0158, J1324 and J1524 systems show excess emission in the residual spectra after subtraction of the template.
}
 

\end{abstract}

   
%section 01           
\section{Introduction} \label{sec:intro}
The EWs are contact binaries in which both components are dwarf stars with spectral class ranging from F to K. They show almost same size primary and secondary minima \citep{1941ApJ....93..133K, 1967AJ.....72S.309L, 1968ApJ...151.1123L}. Both components fill their Roche lobes and show strong tidal interaction. The contact binary systems are quite common. Out of the $\sim$2,083,548 confirmed variable stars in the VSX catalog \citep{2006SASS...25...47W}, almost 25$\%$ are reported as contact binaries (CBs) and $\sim$19\% as EWs. As both the components have a common convective envelope around them, change and/or loss in mass and angular momentum can take place \citep{1967AJ.....72S.309L, 1968ApJ...151.1123L, 1989A&A...220..128V, 2001MNRAS.328..914Q}. An orbital period change can be associated with mass transfer, angular momentum loss due to magnetic activities, or a light-time effect (LITE) because of the presence of third component. Actually, angular momentum loss itself is the leading mechanism behind the formation of EWs from small period detached eclipsing binaries (DEBs). The evolution process is directed by the expansion of the most evolved component accompanied by mass transfer and angular momentum loss due to stellar wind, a third body and/or mass loss. So, the DEBs can evolve to EWs via angular momentum loss (due to magnetic activies), third components (via Kozai cycles) or an evolutionary expansion of its components \citep{1962AJ.....67..591K, 1989A&A...220..128V, 2004MNRAS.355.1383L, 2007ApJ...662..596L, 2008MNRAS.389.1722E}. Uneven sized maxima are very common in photometric light curves (LCs) of EWs. Cool and/or hot spots are believed to be the reason behind this asymmetry of their LCs. Many EWs show excess or filled in emission in their spectra due to magnetic activities \citep{1984BAAS...16..893B, 1984AJ.....89..549H, 1995A&AS..114..287M}. Due to the small separation of the components and the corresponding short orbital periods, these systems can be easily observed with ground-based telescopes on a short timescale. The EWs are also appropriate targets to study many other processes like the O'Connell effect \citep{1951MNRAS.111..642O}, energy exchange between components, magnetic activity, orbital period changing mechanisms, thermal relaxation oscillation (TRO) etc. Photometric and spectroscopic data are essential for an accurate determination of physical parameters of EWs.

In the present work, five EWs are analyzed and their absolute parameters, rate of orbital period change, mass-transfer, emission properties are studied. The system J1524 was first identified as a contact binary by the ROTSE survey \citep{2000AJ....119.1901A}. The other systems were classified as EWs by the Catalina Surveys Periodic Variable Star Catalog which provides the average V-band magnitude, amplitude of variation and period of periodic variables. The first data release of the Catalina Surveys identified $\sim47,000$ periodic variables among 5.4 million candidate stars within a 20,000 $deg^2$ region of sky \citep{2014ApJS..213....9D}. The objects J015818.6+260247 (J0158b), J073248.4+405538 (J0732), J101330.8+494846 (J1013), J132439.8+130747 (J1324) and J152450.7+245943 (J1524)
are EWs, with an average V-band magnitude ranging from 13.5 to 14.5 and an orbital period ranging from 0.244 to 0.286 day in the Catalina Real-time Transient Survey (CRTS) Catalog. The basic information about the targets is listed in Table~\ref{tar_info}.

%Table 01
\begin{table}[ht]
\caption{Basic information about the sources taken from different surveys.}   
%\centering 
\begin{center}             
\label{tar_info}
\scriptsize         
%\begin{tabular}{l c c c c c c c c}    
\begin{tabular}{p{.25in}p{.41in}p{.41in}p{0.41in}p{0.2in}p{0.2in}p{0.2in}p{0.25in}}
\hline
Source& RA         &DEC        & Period   & V      & $A_{V}$ & J-K    & Parallax\\
      & (J2000)    &(J2000)    & (day)    & (mag)  & (mag)   & (mag)  & (mas)   \\
%      & hh:mm:ss   & dd:mm:ss  &          &       &
      &            &           & $^{(1)}$ &$^{(1)}$& $^{(2)}$&$^{(3)}$& $^{(4)}$\\
\hline
J0158b& 01:58:18.6 & +26:02:47 & 0.263009 & 13.60 & 0.240 & 0.305 & 2.0283 \\
J0732 & 07:32:48.4 & +40:55:38 & 0.286025 & 14.61 & 0.151 & 0.668 & 1.0452 \\
J1013 & 10:13:30.8 & +49:48:46 & 0.250206 & 13.72 & 0.019 & 0.494 & 2.5813 \\
J1324 & 13:24:39.8 & +13:07:47 & 0.266082 & 13.59 & 0.060 & 0.494 & 1.9923 \\
J1524 & 15:24:50.7 & +24:59:43 & 0.244760 & 13.67 & 0.114 & 0.593 & 2.1804 \\ 
\hline 
\end{tabular}

\end{center} 
%\begin{tablenotes}
%\scriptsize
%\item References. $^{(1)}$ \citealt{2014ApJS..213....9D}; $^{(2)}$ \citealt{2011ApJ...737..103S}; $^{(3)}$ \citealt{2006AJ....131.1163S} (2MASS survey); $^{(4)}$ \citealt{2020arXiv201201533G} (GAIA).
%\end{tablenotes}

{\raggedright References. $^{(1)}$ \citealt{2014ApJS..213....9D}; $^{(2)}$ \citealt{2011ApJ...737..103S}; $^{(3)}$ \citealt{2006AJ....131.1163S} (2MASS survey); $^{(4)}$ \citealt{2020arXiv201201533G} (GAIA). \par}
\end{table}

%
%section 02
\section{Observations}\label{Data}
\subsection{Photometry}\label{Photo}
For the photometric observations, we used the 1.3\,m Devasthal Fast Optical Telescope (DFOT) and the 1.04\,m Sampurnanand Telescope (ST), operated by the Aryabhatta Research Institute of Observational Sciences (ARIES). The DFOT has a 2k$\times$2k CCD with a $\sim$18$^{\arcmin}\times$18$^{\arcmin}$ field of view (FoV). The DFOT CCD is operated with a gain of 2\,e$^{-}$\,ADU$^{-1}$ and readout noise of 7.5\,$e^{-}$. The ST is equipped with a 4k$\times$4k CCD with a FoV of $\sim$15$^{\arcmin}\times$15$^{\arcmin}$. The ST CCD is used in 4$\times$4 binning mode with a gain of 3\,e$^{-}$\,ADU$^{-1}$ and readout noise of 10\,e$^{-}$. Table~\ref{log_phot} shows the observation log for DFOT and ST photometric observations. For the pre-processing of raw images, IRAF\footnote{IRAF is distributed by the National Optical Astronomy Observatory, which is operated by the Association of Universities for Research in Astronomy under cooperative agreement with the National Science Foundation.} routines were used. The standard procedure of bias subtraction, flat-fielding, cosmic ray removal was followed. The instrumental magnitudes were determined through aperture photometry using DAOPHOT \citep{1992ASPC...25..297S}. The differential LCs were generated after choosing suitable comparison stars close to the target stars in the field.

We also used the photometric observations by the Transiting Exoplanet Survey Satellite (TESS). The TESS was launched in 2018 in search of exoplanets with sizes in the range between earth-like planets and gas giants. It is an all-sky survey that uses four cameras with a FoV of 24$^{\circ}\times$24$^{\circ}$ each as back-end instruments. These cameras are combined to observe sectors of 24$^{\circ}\times$96$^{\circ}$. For the prime mission (first two years), the region of the sky with an ecliptic lattitude $|b|$\,$>$\,6$^{\circ}$ was divided into 26 different sectors to observe each of them for 27 days \citep{2015JATIS...1a4003R}. In this period, TESS observed 200,000 pre-selected targets with a 2-min candence while all the objects on the full frame images (FFIs) have an observation cadence of 30\,min. TESS is now in the extended phase where large parts of the both ecliptic hemispheres are being re-observed but also attention is given to the region near the ecliptic equator. In this phase, the candence of the FFIs has been increased to 10\,min. The TESS data are publicly available at the the Barbara A. Mikulski Archive for Space Telescopes (MAST) portal \footnote{https://mast.stsci.edu}. The data products are available in form of FFIs, target pixel files (TP; time-series for calibrated pixels saved at 2-min. candence) and light curve files (LC; flux time series determined from the TP files). In the LC files, the simple aperture photometry flux (SAP$\_$FLUX) and detrended flux (PDCSAP$\_$FLUX) is available. We used the PDCSAP$\_$FLUX during analysis as it was derived after removing effects of nearby stars and other systematic trends. 

The photometric data from other surveys was also used for period analysis but most of the times of minima (TOMs) were derived from SuperWASP (Wide Angle Search for Planets) data. The SuperWASP project is an ultra-wide-angle photometric survey observing north and south hemispheres of the sky from the sites of  La Palma, Canary Islands and South African Astronomical Observatory (SAAO) Sutherland. Each site covers almost 500 $deg^2$ part of the sky with help of eight cameras. The aperture size of each camera is 11.1 cm and uses 2K$\times$2K EEV CCD. The FoV of individual camera is around 7.8$\times$7.8 $deg^2$ with plate scale of 13.7 arcsec per pixel \citep{2006Ap&SS.304..253P}. The individual object co-ordinates or ID can be used for query on SuperWASP public archive \footnote{https://wasp.cerit-sc.cz/form}. The archive contains time series data of almost 18 million target observed during SuperWASP survey.
%Table 02
\begin{table}[]
\caption{The 1.3\,m DFOT and 1.04\,m ST observation log. The '*' at the end of rows indicate the 1.04\,m ST data.}
\begin{center} 
\label{log_phot}
\scriptsize
\begin{tabular}{p{.55in}p{.5in}p{0.5in}p{0.5in}p{0.55in}p{0.1in}}
\hline
Date of   & Start Jd  &  End Jd   & Total      & Exposure       & Obs.      \\
obs.      &           &           & frames     & time           & time      \\
          & (2450000+)&(2450000+) &            & (sec)          & (hrs)     \\
\hline                         
          &           &           &V, $R_{c}$, $I_{c}$&                &           \\
          &           &           &J0158b      &                &           \\
2018-10-11& 8403.4153 & 8403.4265 & 02, 02, 02 & 180,120,80     &   0.27    \\
2018-11-20& 8443.0701 & 8443.3544 & 51, 51, 50 & 120, 60,50     &   6.82    \\
2018-12-27& 8480.1347 & 8480.1715 & 06, 06, 06 & 180,120,80     &   0.88    \\
2019-10-14& 8771.1721 & 8771.4613 & 84, 84, 84 & 40, 25,20      &   6.94    \\
2020-10-14& 9137.3781 & 9137.4570 & 120, -, 140& 20, - ,20      &   6.94    \\
2020-10-15& 9138.3428 & 9138.4379 & 100, 100, -& 30, 30, -      &   2.28    \\
\hline
          &           &           &J0732       &                &           \\
2021-02-25& 9271.0616 & 9271.3119 & 38, 38, 38 & 120,120,120    &   6.01    \\
2021-02-27& 9273.0622 & 9273.3219 & 40, 40, 40 & 120,120,120    &   6.23    \\
2021-03-01& 9275.0603 & 9275.2778 & 75, 75, 75 & 60, 60, 60     &   5.22    \\
2021-03-26& 9300.0706 & 9300.2491 & 26, 28, 20 & 60, 60, 60     &   4.28*   \\
2021-03-28& 9302.1333 & 9302.2521 & 15, 19, 08 & 90, 90, 90     &   2.85*   \\
2021-03-29& 9303.0723 & 9303.1853 & 37, - , -  & 120, - , -     &   2.71*   \\
2021-04-06& 9311.0651 & 9311.1133 & 12, 12, 12 & 60, 60, 60     &   1.16*   \\
2021-04-08& 9313.0745 & 9313.1281 &  - , - , 27&  - , - , 60    &   1.29*   \\
\hline
          &           &           &J1324       &                &           \\
2021-01-29& 9244.3259 & 9244.5367 & 38, 37, 37 & 120, 80, 80    &   5.06    \\
2021-01-30& 9245.3338 & 9245.5395 & 37, 36, 36 & 120, 80, 80    &   4.94    \\
2021-02-11& 9257.4633 & 9257.5193 & 30, - , -  & 150, - , -     &   1.34    \\
2021-02-17& 9263.4822 & 9263.5245 & 30, - , -  & 120, - , -     &   1.01    \\
2021-03-26& 9300.2607 & 9300.5029 & 73, 73, 73 & 60, 60, 60     &   5.81*    \\
\hline
          &           &           &J1013       &                &           \\
          &           &           & B, V,$I_{c}$&                &           \\
2019-03-19& 8562.2631 & 8562.2882 &  - , 05, 05&  - ,120, 60   &   0.60    \\
2019-03-20& 8563.2192 & 8563.2279 &  - , 02, 02&  - ,120, 60   &   0.21    \\
2020-04-06& 8946.1182 & 8946.2560 & 19, 19, 21 & 150,120, 60   &   3.31    \\
2020-04-08& 8948.0701 & 8948.2594 & 27, 27, 27 & 150,120, 60   &   4.54    \\
2020-04-19& 8959.1740 & 8959.2160 & 06, 07, 06 & 150,120, 60   &   1.01    \\
2020-04-22& 8962.1405 & 8962.2510 & 22, 22, 22 & 90, 60, 30     &   2.65    \\
2020-11-07& 9161.4246 & 9161.5170 &  - , 81, - &  - , 60, -     &   2.22    \\
2020-11-08& 9162.4754 & 9162.4975 &  - , 30, - &  - , 60, -     &   0.53    \\
2020-11-14& 9168.4224 & 9168.4982 &  - , 100, -&  - , 60, -     &   1.82    \\
2021-03-21& 9295.1523 & 9295.1905 & 35, - , -  & 90, - , -      &   0.92    \\
2021-03-27& 9301.1317 & 9301.3740 & 46, 36, 47 & 150, 90, 60    &   5.81*    \\
2021-03-28& 9302.2724 & 9302.3657 & 51, - , -  & 150, - , -     &   2.24*    \\
2021-03-29& 9303.2060 & 9303.3640 & 39, 40, 39 & 60, 60, 60     &   3.79*    \\
2021-03-30& 9304.0771 & 9304.3503 & 49, 46, 48 & 180, 90, 60    &   6.56*    \\
\hline
          &           &           &J1524       &                &           \\
2018-05-25& 8264.1858 & 8264.3735 & 15, 15, 16 & 300,180, 60   &   4.50    \\
2018-05-26& 8265.1056 & 8265.1889 & 09, 09, 09 & 300,150, 60   &   2.00    \\
2019-03-15& 8558.4717 & 8558.5088 & 06, 06, 06 & 200,150, 60   &   0.89    \\
2019-03-19& 8562.4843 & 8562.5079 & 04, 03, 03 & 200,150, 60   &   0.57    \\
2019-03-21& 8564.2599 & 8564.3224 & 08, 09, 09 & 200,150, 60   &   1.50    \\
2019-03-28& 8571.3943 & 8571.4934 & 18, 18, 18 & 120, 60, 30    &   2.38    \\
2019-03-31& 8574.4320 & 8574.4641 & 05, 05, 05 & 150,100, 60   &   0.77    \\
2019-05-18& 8622.3634 & 8622.4269 & 14, 07, 07 & 150,100, 80   &   1.53    \\
2019-05-19& 8623.2524 & 8623.3421 & 13, 13, 13 & 150,100, 80   &   2.15    \\
2020-03-28& 8937.4586 & 8937.4969 & 08, 08, 11 & 60, 40, 30     &   0.92    \\
2020-03-29& 8938.3352 & 8938.4886 & 35, 35, 35 & 60, 40, 30     &   3.68    \\
2020-04-19& 8959.4248 & 8959.4763 & 08, 08, 09 & 120, 90, 60    &   1.23    \\
2020-04-21& 8961.2938 & 8961.4712 & 16, 16, 16 & 90, 60, 30     &   4.26    \\
2021-03-27& 9301.4126 & 9301.5016 &  - , 114, -&  - , 60, -     &   2.13    \\
2021-03-28& 9302.3826 & 9302.4914 & 59, - , -  & 150, - , -     &   2.61    \\
\hline                                                    
\end{tabular}  
\end{center}      
\end{table}

%
% Figure 01
\begin{figure*}[!ht]
\begin{center}
\subfigure{\includegraphics[width=18cm,height=5.4cm]{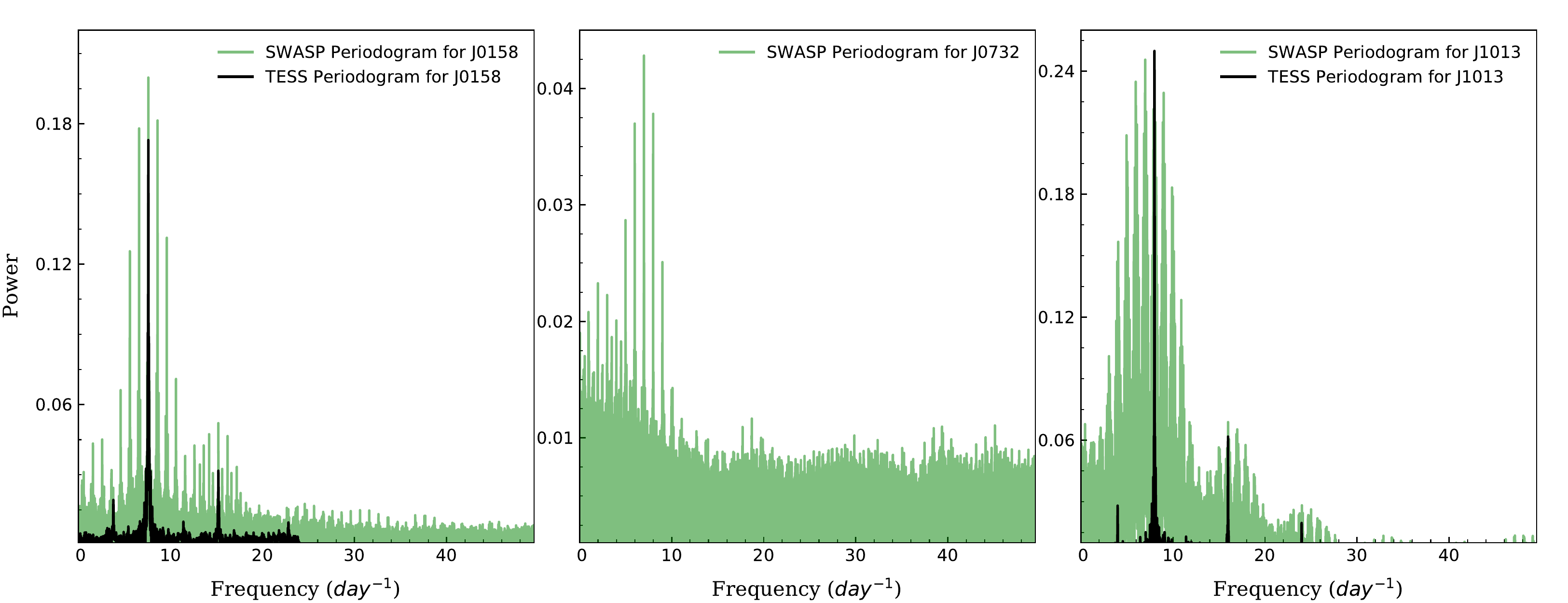}}
\subfigure{\includegraphics[width=12cm,height=5.4cm]{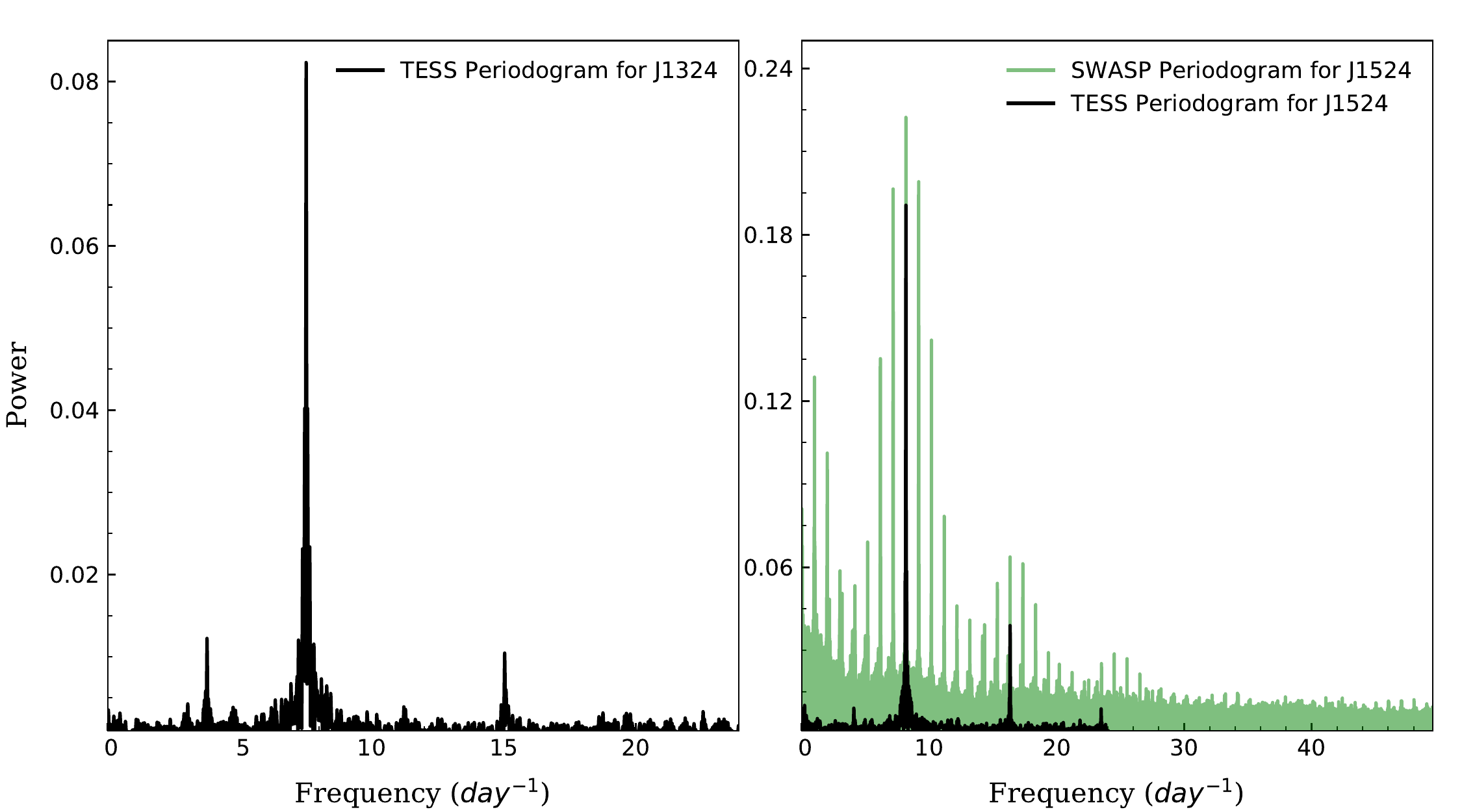}}
\caption{
Power spectra obtained by applying the PERIOD04 software to the SuperWASP (light green) and TESS (black) data of the targets. The object names are given at the top right corner of each plot.
}
\label{periodo}
\end{center}
\end{figure*}
\subsection{Spectroscopy}\label{Spec}
The Large sky Area Multi-Object Fibre Spectroscopic Telescope (LAMOST) survey provides a huge collection of low-resolution spectra (R$\sim$1800) of stars, galaxies, quasars and other unknown objects. The multi-fiber design allows to observe 4000 targets simultaneously \citep{2015RAA....15.1095L}. Since the start of the second phase of the regular survey in September 2018, the medium-resolution spectra (R$\sim$7500) are also gathered. The second version of the 4th data release of the LAMOST survey includes 1,551,394 star spectra, 39,498 galaxy spectra and 13,954 quasar spectra. The available low-resolution data of targets were downloaded from the LAMOST DR4 v2 website \footnote{http://dr4.lamost.org/}. Table~\ref{tar_lamost} give an overview of these observations and the published values of the spectral subclass and stellar parameters (effective temperature \teff, surface grativy \logg, and metallicity \feh), derived with the LAMOST Stellar Parameter Pipeline (LASP). The LASP uses ULySS package for estimating temperature and other stellar parameters from the observed spectra. The stellar spectra are compared with similar resolution model spectra to minimize the $\chi2$ \citep{2015RAA....15.1095L}. SNR$_i$ denotes the signal-to-noise ratio in the Slaon $i$-band. All targets are classified as main-sequence late-type objects.

%Table 03
\begin{table}[!ht]
\caption{
%Parameters of targets from the LAMOST data
Parameters of targets as listed in the catalogue stellar parameters of low-resolution LAMOST spectra (DR6 v2). The errors in units of the last decimal are given between parentheses.
}             
\label{tar_lamost}
\begin{center}  
\scriptsize       
\begin{tabular}{p{.3in}p{.52in}p{.38in}p{0.25in}p{0.35in}p{0.45in}p{0.1in}}
\hline   
Targets & Date        & \teff     & Sub & \logg      & \feh   & SNR$_i$\\
        &             &  (K)      &class&            & (dex)  &        \\
\hline
J0158b  & 31-10-2013  & 5599(26)  & G7  & 4.35(4)  & -0.34(3) &    144 \\ %14:57 UT
	& 31-10-2013  & 5316(33)  & G7  & 3.95(6)  & -0.50(3) &    138 \\ %15:54 UT
	& 08-12-2014  & 5562(17)  & G7  & 4.28(3)  & -0.35(2) &    274 \\ %13:34 UT
\hline                                                                                     
J0732   & 07-03-2015  & 5494(183) & G7  & 4.16(19) & -0.27(19)&     77 \\ %12:04 UT
\hline                                                                                     
J1013   & 23-01-2013  & 4926(140) & K3  & 4.57(18) & -0.03(13)&     50 \\ %18:11 UT
        & 05-02-2015  & ---       & --- & ---      & ---      &     12 \\ %18:31 UT
\hline                                                                                     
J1324   & 18-01-2012  & 4997(58)  & K1  & 4.60(10) & -0.32(6) &    103 \\ %22:17 UT
\hline                                                                                      
J1524   & 12-03-2016  & 5198(22)  & G7  & 4.00(4)  & -0.71(2) &    173 \\ %20:34 UT
\hline
\end{tabular}
\end{center} 
\end{table}

%
% Figure 02
\begin{figure*}[!ht]
\begin{center}
\subfigure{\includegraphics[width=17cm,height=5.2cm]{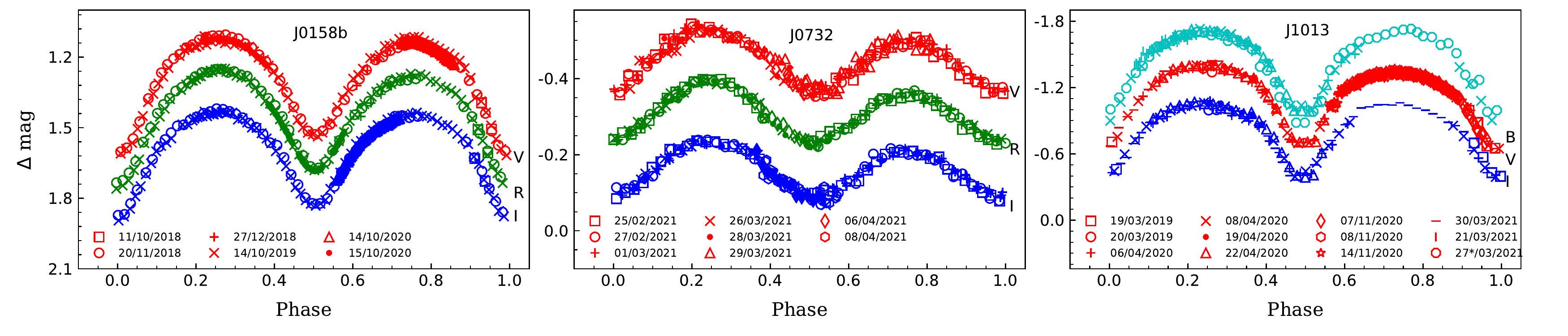}}\vspace{-0.5cm}
\subfigure{\includegraphics[width=12cm,height=5.2cm]{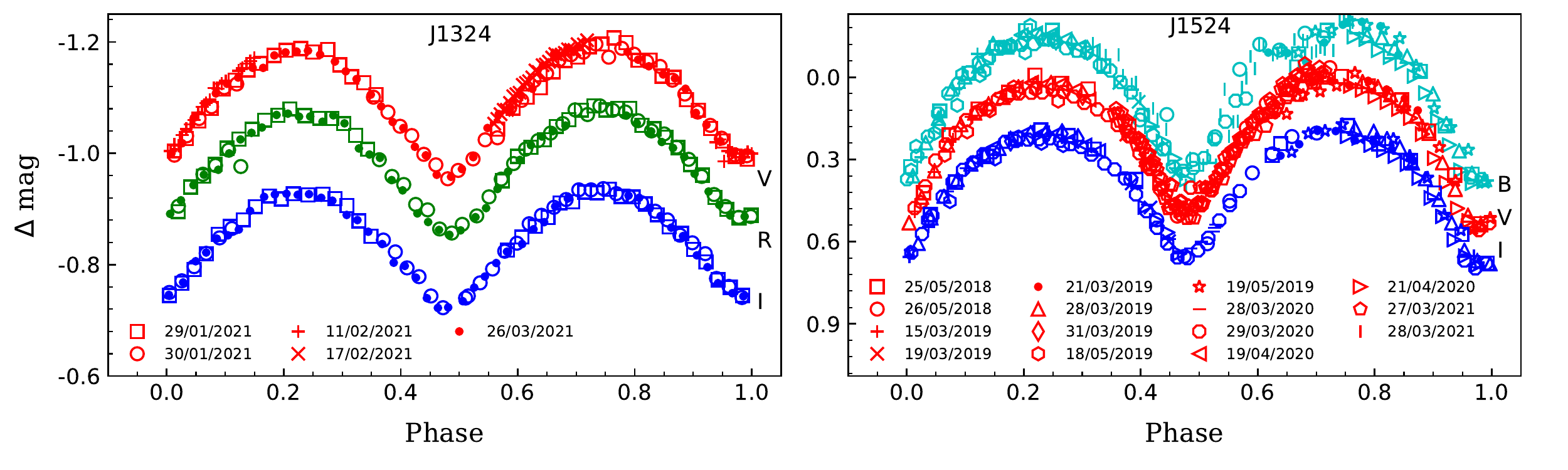}}
\caption{The multi-band light curves of the five targets observed using 1.3\,m DFOT and 1.04\,m ST. The different colors represent different bands also mentioned in the right side of each plot. The different symbols correspond to different dates of observations.}
\label{lc_obs}
\end{center}
\end{figure*}
%
%section 03
\section{Orbital period and orbital period change}\label{orpe}
We analyzed the time series for these objects in order to detect any secular changes in their orbital period \porb. The period analysis is an effective way for understanding processes like mass loss/transfer, long term activity cycles, the influence of a third component and the dynamical evolution of eclipsing binaries. Accurate \porb values were determined by applying the PERIOD04 software \citep{2004IAUS..224..786L, 2005CoAst.146...53L} to the SuperWASP and TESS data of the targets. This software can handle time series with large gaps and multiperiodic signals. SuperWASP data are available for J0158b, J0732, J1013 and J1524 while the TESS mission observed all the targets except J0732. These targets were also observed in many other surveys but due to the large time span and better cadence of SuperWASP and TESS data, only these surveys data were used to determine period. The power spectra that were used for the period search are shown in Figure~\ref{periodo}. The SuperWASP data power spectra for J0158b, J0732, J1013 and J1524 show peak at frequencies 7.604314(4), 6.99249(4), 7.9934(2) and 8.17127(8) $day^{-1}$, respectively. The power spectra derived from TESS data show peak at frequencies 7.6036(2), 7.9932(2), 7.5163(2) and 8.1721(1) for J0158b, J1013, J1324 and J1524, respectively. The primary and secondary minima are almost indistinguisable due to nearly equal depths, so, the periods ($1/freq_{peak}$) given by periodogram are half of the actual period i.e. $ 2/freq_{peak}$. The periods are determined as 0.263033(7), 0.250213(6), 0.266088(7) and 0.244735(3) day for J0158b, J1013, J1324 and J1524, respectively, from the TESS photometric time series. The periods from SuperWASP data are found to be 0.2630086(1), 0.286021(2), 0.250206(6) and 0.244759(2) for J0158b, J0732, J1013 and J1524, respectively.  The periods obtained from SuperWASP and TESS are very close to the previously reported values in \cite{2014yCat..22130009D}. The phase folded observed LCs derived from estimated periods in different photometric bands are shown in Figure~\ref{lc_obs}.

The change in orbital period with time is investigated by studying the variations in time of minima (TOMs) at different epochs. 
For this, we used data from different surveys such as the CRTS, SuperWASP \citep{2010A&A...520L..10B}, Zwicky Transient Facility (ZTF; \cite{2020MNRAS.499.5782O, 2014htu..conf...27B}), Palomar Transient Factory (PTF; \cite{2012PASP..124..854O}), TESS, and 1.3\,m DFOT observations. For SuperWASP, there was a small shift in the observed magnitudes corresponding to different SuperWASP cameras on some specific days of observations. We therefore separated all data points according to the camera ID e.g. all data points observed using camera-101 in one file, camera-102 in other file and so on. Each night LCs were plotted using these data files prepared on the basis of camera IDs. We visually analyzed the LC of each individual night and only selected the good quality LCs for the TOM determination (LCs with less scattering and covering at least $>50\%$ part of the phase around one of the minima). In some cases, one of the minima was observed but number of data points was small around minima region and not enough for TOM determination using parabola fitting. In these cases, data from consecutive days were combined together to get a combined LC with good number of data points in minima region. The TOM was determined by parabola fitting to the data around primary or secondary minima. A similar procedure was used to determine TOM for data collected from other surveys. In the following subsections, we describe the orbital period change analysis based on these TOMs for each target individually.

%----------------------------------------
\subsection{J0158b}\label{J0158b_per_stu}
%----------------------------------------

% Figure 3
\begin{figure*}[!ht]
\label{oc_0158_1013}
\begin{center}
\subfigure{\includegraphics[width=5.75cm,height=6.2cm]{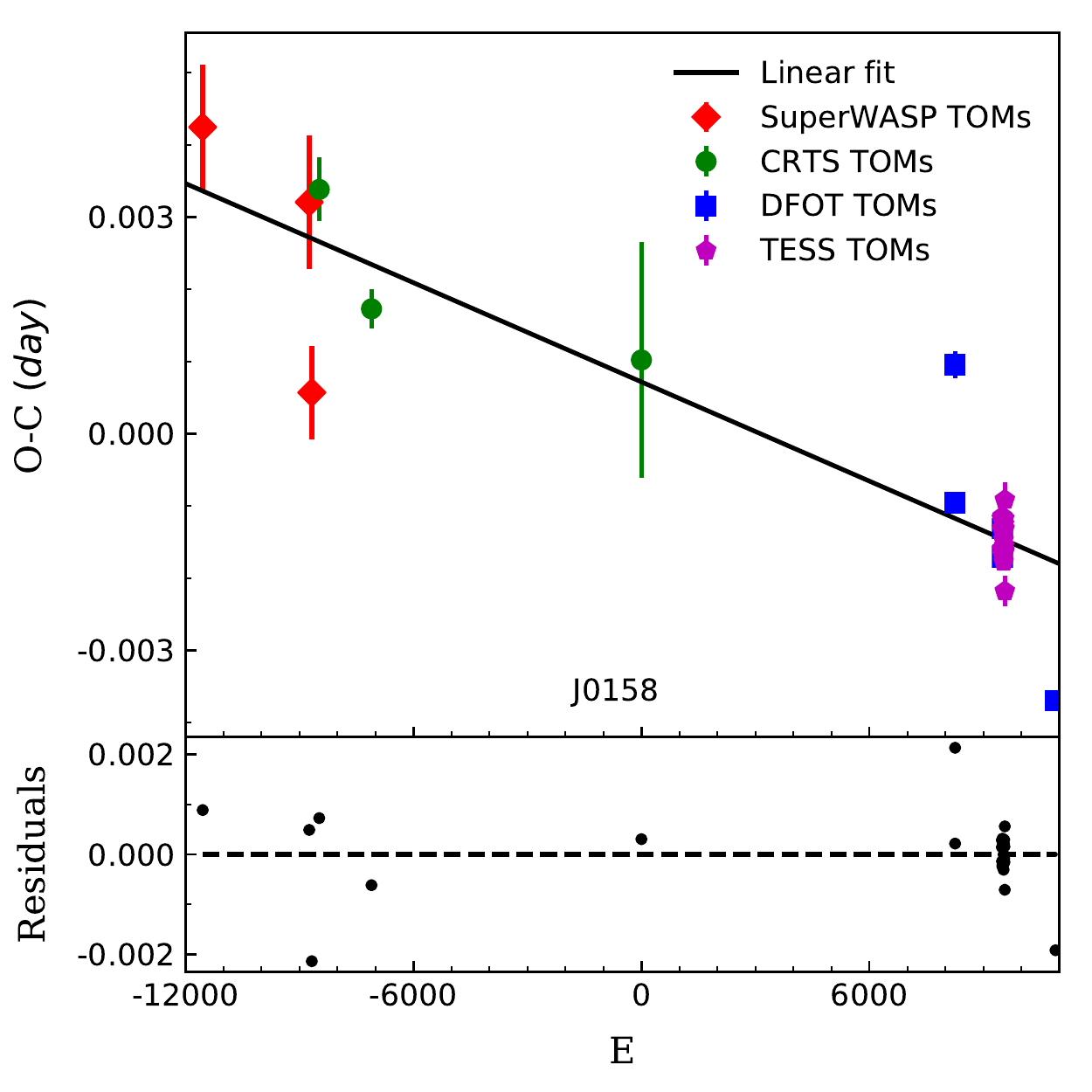}}
\subfigure{\includegraphics[width=5.75cm,height=6.2cm]{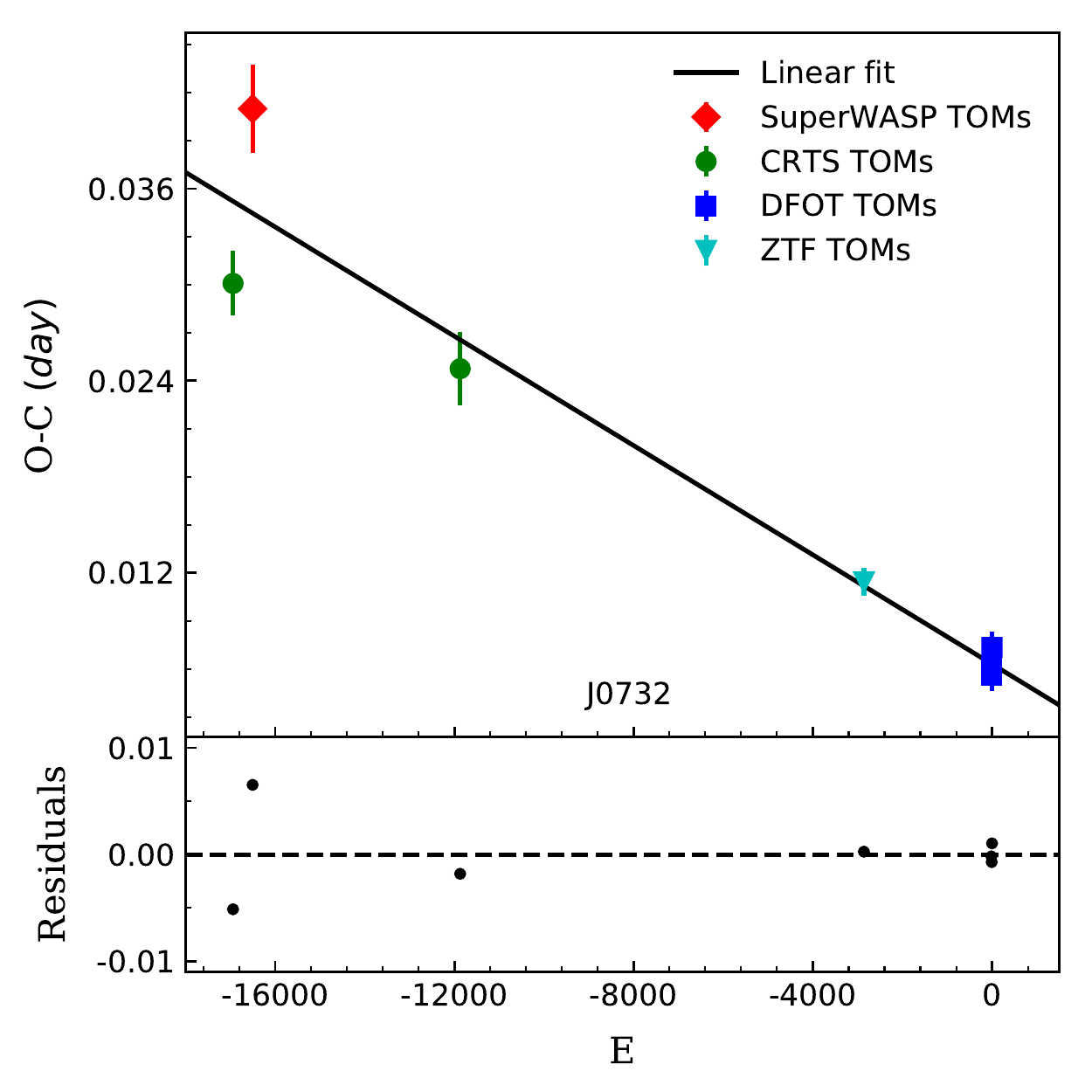}}
\subfigure{\includegraphics[width=5.75cm,height=6.2cm]{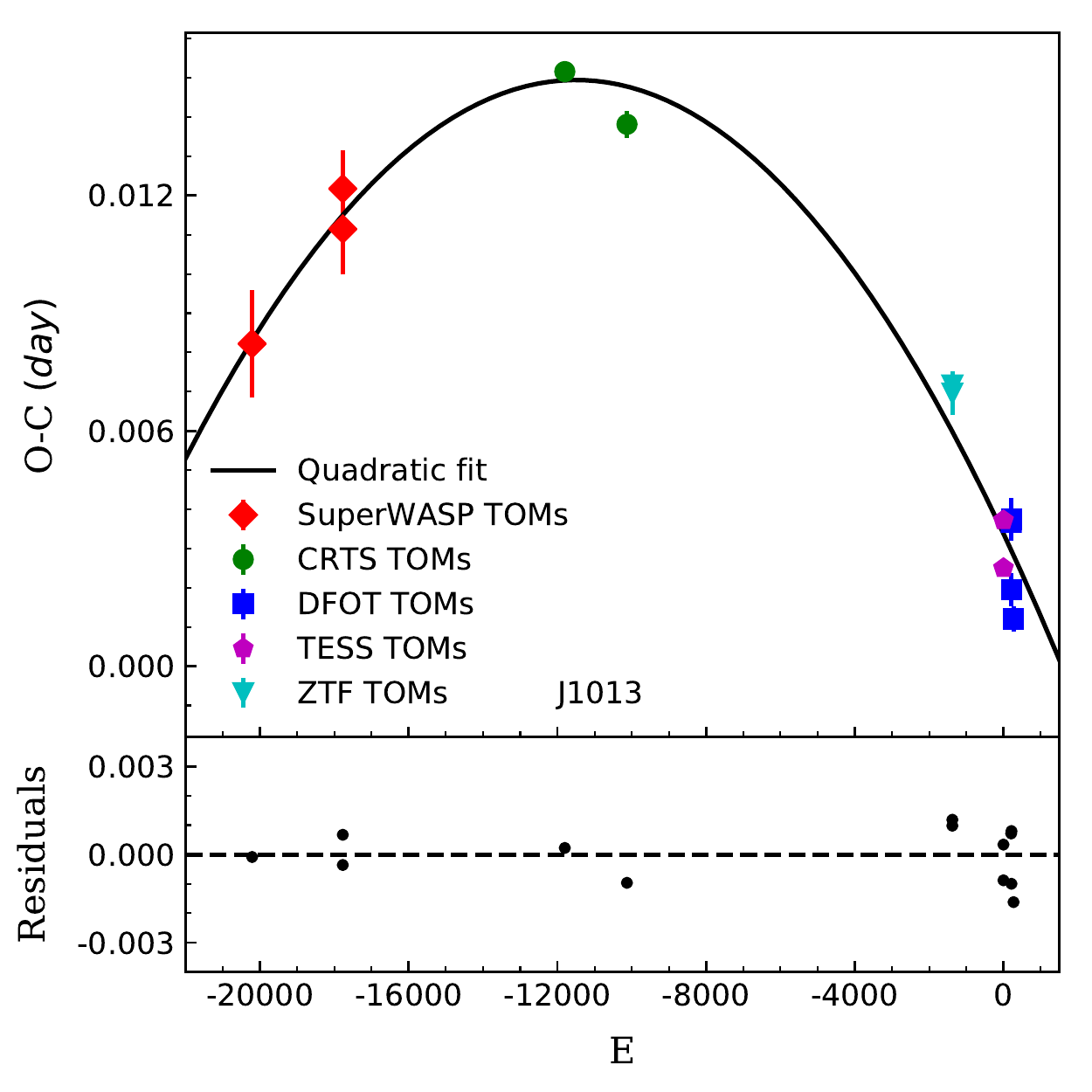}}
\subfigure{\includegraphics[width=5.75cm,height=6.2cm]{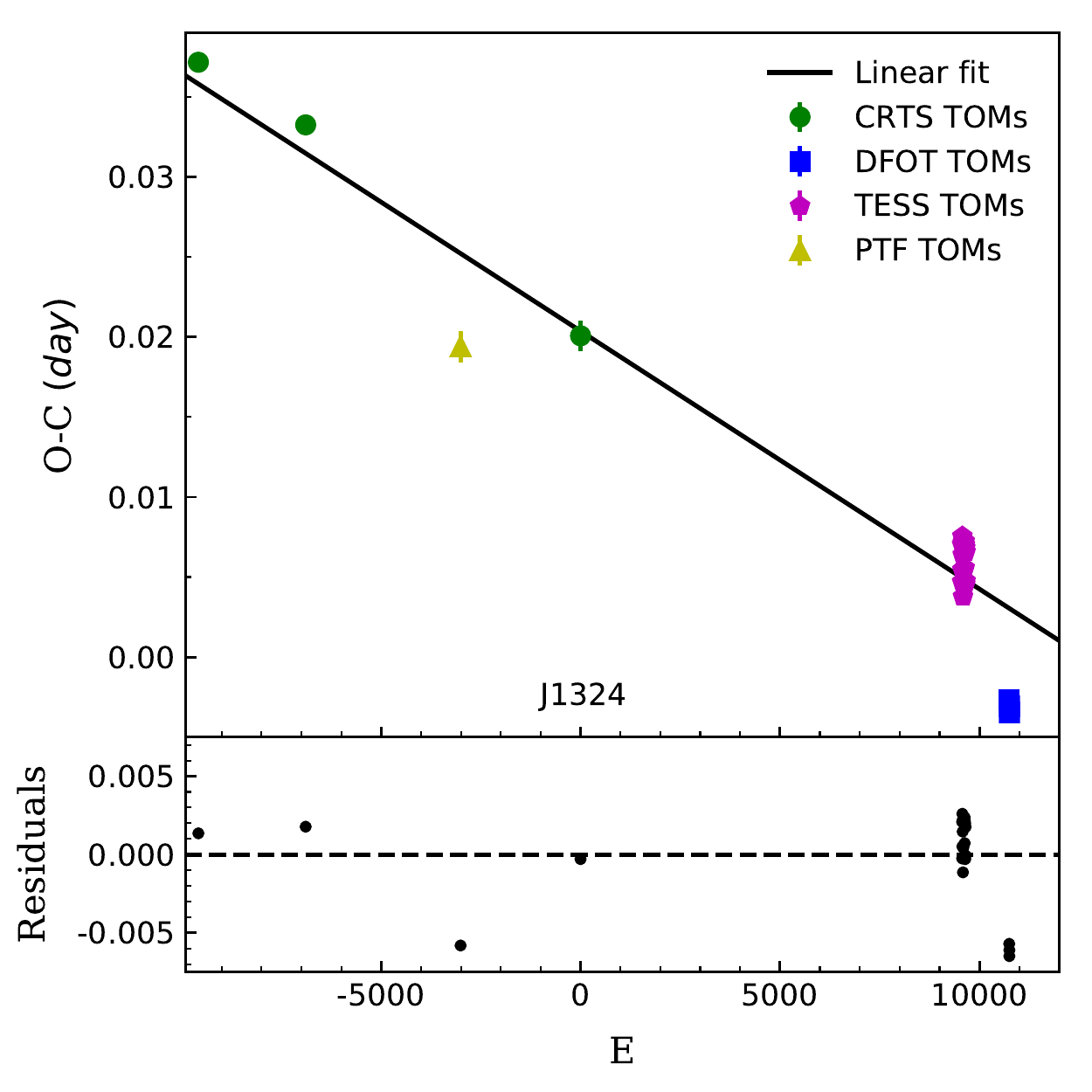}}
\subfigure{\includegraphics[width=5.75cm,height=6.2cm]{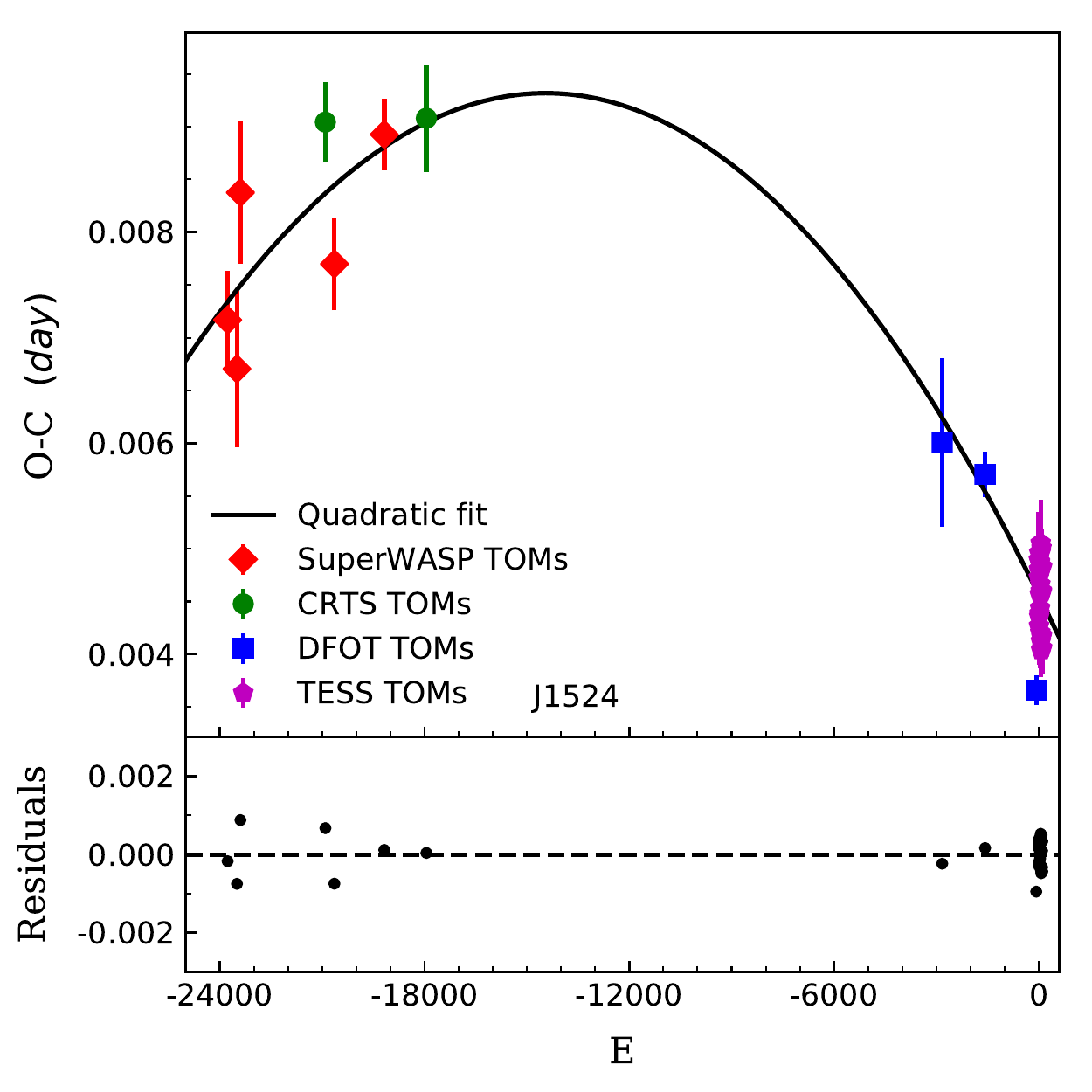}}
\caption{ The O-C diagrams of the targets with a linear/quadratic fit. The x-axis corresponds to the orbital cycle number and y-axis represents O-C difference of TOMs. The residuals of the fit are shown in lower subpanels. The target name is mentioned in each plot. 
}
\end{center}
\end{figure*}
The three TOMs were determined using SuperWASP photometric observations. The observations by CAM-101 from JD 2453229 to 2453237, CAM-141 from JD 2453970 to 2453972 and CAM-142 from JD 2453979 to 2453998, were combined together to calculate two primary and one secondary TOMs. A total of three secondary minima were calculated from CRTS data (data of 80 to 100 days was combined for generating LCs). We also determined 16 TOMs from TESS data and five TOMs from DFOT data. The O-C was calculated using an orbital period of 0.263009 day. 
The linear ephemeris was computed by line fitting to the orbital cycle-TOM curve which can be expressed as :
\begin{equation}
\label{li_58}
\begin{aligned}
HJD_{o}&=6271.233769(\pm0.000198)+0.26300877\\
&(\pm0.00000002)\times E
\end{aligned}
\end{equation} 
Here, $HJD_{o}$ represents (HJD$_{TOM}$-2,450,000) at primary minimum of orbital cycle number E.
The top left panel of Figure~\ref{oc_0158_1013} shows the best fitted line to the E v/s (O-C) plot. The lower part shows the residuals of the fit. The fitted line can be represented by:
\begin{equation}
\label{li_58b_oc}
\begin{aligned}
O-C &=0.0007(\pm0.0002)-2.2901(\pm0.2148) \times10^{-7} \times E
\end{aligned}
\end{equation}
The TOMs from SuperWASP and CRTS show larger error bars as compared to those from TESS and DFOT data TOMs because of the quality of photometric observations. From the top left panel of Figure~\ref{oc_0158_1013}, we conclude that there is no evidence for a change of the \porb\ value of J0158b in the time span of the photometric observations that were used in the present study.
% Table 04
\begin{table}[!ht]
\caption{Eclipse minima timings for J0158b, J0732, J1013, J1324 and J1524.}
\label{OC_info}
\begin{center} 
\scriptsize
%\begin{tabular}{l c c c c c c c}   
\begin{tabular}{p{.27in}p{.8in}p{0.15in}p{0.35in}p{0.5in}p{0.5in}p{0.05in}}
\hline
ID     & $HJD_{o}$       &Min& Cycle & $(O-C)_{1}$ & $(O-C)_{2}$ & Ref \\
       &(2450000+)       &   &       &  (days)     & (days)      &     \\
\hline
J0158b& 3233.220340(864) & p & -11551   &  0.004246 &  0.000884 & 1 \\
J0158b& 3971.091050(925) & s &  -8745.5 &  0.003207 &  0.000487 & 1 \\
J0158b& ---              & s &  ---     &  ---      &  ---      & - \\
J0158b& ---              & s &  ---     &  ---      &  ---      & - \\
J0732 & 8457.325452(893) & s &  -2859.5 &  0.011446 &  0.000265 & 5 \\
J0732 & 9275.209839(968) & p &      0.0 &  0.007345 &  0.001044 & 3 \\
J0732 & ---              & s &  ---     &  ---      &  ---      & - \\
J0732 & ---              & s &  ---     &  ---      &  ---      & - \\
J1013 & 8550.278425(558) & s &  -1375.0 &  0.006963 &  0.000981 & 5 \\
J1013 & 8550.403730(269) & p &  -1374.5 &  0.007166 &  0.001184 & 5 \\
J1013 & ---              & s &  ---     &  ---      &  ---      & - \\
J1013 & ---              & s &  ---     &  ---      &  ---      & - \\
J1324 & 3835.832758(471) & p &  -9580.5 &  0.037157 &  0.001352 & 2 \\
J1324 & 4551.589427(623) & p &  -6890.5 &  0.033246 &  0.001776 & 2 \\
J1324 & ---              & s &  ---     &  ---      &  ---      & - \\
J1324 & ---              & s &  ---     &  ---      &  ---      & - \\
J1524 & 8956.458941(457) & p &      0.0 &  0.004895 &  0.000333 & 4 \\
J1524 & 8970.288040(362) & p &     56.5 &  0.005054 &  0.000529 & 4 \\
J1524 & ---              & s &  ---     &  ---      &  ---      & - \\
J1524 & ---              & s &  ---     &  ---      &  ---      & - \\
\hline                  
\end{tabular}
\end{center} 
{\raggedright Here [1], [2], [3], [4], [5] and [6] show TOMs obtained by SuperWASP, CRTS, DFOT, TESS, ZTF and PTF. This is only sample table and the full table is only available in the online version of the paper. \par}
\end{table}

%--------------------------------------
\subsection{J0732}\label{J0732_per_stu}
%--------------------------------------

J0732 is the only system in our sample for which no TESS data is available. A total of 7 TOMs (one from CRTS, two from SuperWASP, one from ZTF and three from DFOT data) were evaluated for J0732. The updated ephemeris was found to be:
\begin{equation}
\label{li_07}
\begin{aligned}
HJD_{o}&=9275.20879(\pm0.00199)+0.2860233\\
&(\pm0.0000002)\times E
\end{aligned}
\end{equation} 
Similar to the J0158b system, the $O-C$ variation with E can be represented by a straight line. The fitted line is shown in the top middle panel of Figure~\ref{oc_0158_1013} and defined as:
\begin{equation}
\label{li_07_oc}
\begin{aligned}
O-C &=0.006(\pm0.002)-1.7065(\pm0.1981)\times 10^{-6} \times E
\end{aligned}
\end{equation}
Since this system also shows linearity in the O-C diagram, we can say its period remained constant for the last 13 years.

%--------------------------------------
\subsection{J1013}\label{J1013_per_stu}
%--------------------------------------

For J1013, 13 TOMs were calculcated (three from SuperWASP, two from CRTS, two from  ZTF, two from TESS, and four using DFOT data). The linear ephemeris for J1013 follows the equation:
\begin{equation}
\label{li_10}
\begin{aligned}
HJD_{o}&=8894.309(\pm0.001)+0.2502052\\
&(\pm0.0000001)\times E
\end{aligned}
\end{equation} 
The updated quadratic ephemeris for J1013 can be represented by:
\begin{equation}
\label{qu_10}
\begin{aligned}
HJD_{o} &=8894.3077(\pm0.0003)+0.2502037(\pm0.0000001)\\
& \times E -8.745(\pm0.852) \times 10^{-11} \times E^{2}
\end{aligned}
\end{equation}
The non-linear variation is obvious in the O-C diagram shown in the bottom left panel of Figure~\ref{oc_0158_1013}. The O-C variations are best fitted with the following quadratic expression:
\begin{equation}
\label{qu_10_oc}
\begin{aligned}
O-C &=0.00338(\pm0.00036)-2.01119(\pm0.15641) \times 10^{-6}\\
& \times E -8.746(\pm0.852) \times 10^{-11} \times E^{2}
\end{aligned}
\end{equation}
The non-linear O-C variations of J1013 are due to a change in its orbital period with a rate of $-2.552(\pm0.249)\times 10^{-7}$ days/year.

%--------------------------------------
\subsection{J1324}\label{J1324_per_stu}
%--------------------------------------

J1324 was not observed in the SuperWASP survey. We determined a total of 26 TOMs (three using CRTS, one using PTF, 19 using TESS and three using DFOT data) for this system. The linear ephemeris is derived as follows:
\begin{equation}
\label{li_13}
\begin{aligned}
HJD_{o}&=6385.01457(\pm0.00096)+0.2660804\\
&(\pm0.0000001)\times E
\end{aligned}
\end{equation} 
The $O-C$ variation with E shows a linear variation (see top right panel of Figure~\ref{oc_0158_1013}).
The fitted line equation is expressed as:
\begin{equation}
\label{li_13_oc}
\begin{aligned}
O-C &=0.0204(\pm0.0009)-1.611598(\pm0.103038)\\
&\times 10^{-6} \times E
\end{aligned}
\end{equation}
As for the systems J0158b and J0732, this system also does not show any noticeable change in \porb.

%--------------------------------------
\subsection{J1524}\label{J1524_per_stu}
%--------------------------------------

In addition to three DFOT TOMs, we determined five using SuperWASP, two using CRTS and 27 using TESS data for J1524. In totality, we used 38 eclipsing timings for analysis of J1524. From the estimated TOMs, the updated linear ephemeris is derived as:
\begin{equation}
\label{li_15}
\begin{aligned}
HJD_{o}&=8956.4587(\pm0.0001)+0.24475983\\
&(\pm0.00000001)\times E
\end{aligned}
\end{equation} 
The least-square solution for quadratic ephemeris is given by:
\begin{equation}
\label{qu_15}
\begin{aligned}
HJD_{o} &=8956.45861(\pm0.00007)+0.24475934(\pm0.00000007)\\
& \times E -2.2773(\pm0.3192) \times 10^{-11} \times E^{2}
\end{aligned}
\end{equation}
The (O-C) diagram with residuals for J1524 is shown in the bottom right panel of Figure~\ref{oc_0158_1013}. A non-linear O-C variation with epochs is clearly noticeable in the diagram. A quadratic fit is drawn for this variation. The following equation represents the non-linear behavior of O-C for J1524 :
\begin{equation}
\label{qu_15_oc}
\begin{aligned}
O-C &=-0.00456(\pm0.00007)-6.5818(\pm0.7008) \times 10^{-7}\\
& \times E -2.2773(\pm0.3192) \times 10^{-11} \times E^{2}
\end{aligned}
\end{equation}
The shape of the O-C curve is similar to a downward parabola as exhibited in case of J1524. The rate of orbital period change is calculated using the Equation~\ref{qu_15_oc}. The system shows period variation of $-6.792(\pm0.952)\times 10^{-8}$ days/year. Some of the TOMs for each target are listed in table~\ref{OC_info} which is a sample table. 
%section 04
%=============================
\section{Light Curve Modeling}\label{Ana}
%=============================

The legacy version of PHOEBE-1.0 (PHysics Of Eclipsing BinariEs) was used for analysis of photometric LCs. The program is based on Wilson-Devinney code \citep{1971ApJ...166..605W}. Due to the release of more advanced version PHOEBE 2 the legacy version of PHOEBE is not actively maintained. Although PHOEBE 2 is more precise having some new features but the legacy version is efficient and more tested than the new version. The legacy version provides a graphical user interface (GUI) and a scripter for fitting and reproducing LCs and radial velocity (RV) curves. The GUI helps to analyze the reproduced LCs after each iteration and the scripter can be used for the analysis of large data sets. The PHOEBE scripter helps in the analysis as well as the statistical tests for the obtained results \citep{2005ApJ...628..426P}. We used the differential corrections minimization method while modeling the targets. Depending upon the Roche lobe geometry (Detached, Semi-detached or Contact binary) of the system, different models can be selected in PHOEBE. Currently, there are eight models available in PHOEBE. We used the "over contact binary not in thermal contact" model for initial estimate of parameters. This model assumes that the components are in geometrical contact but their temperature can be different. In this model, the secondary temperature is independent of the primary temperature. If the system under study is in thermal contact then both the models give same results.

%---------------------------------
\subsection{Effective Temperature}\label{teff}
%---------------------------------

%Table 05
\begin{table}[!ht]
\caption{$T_{eff}$ (in $K$) determined from different empirical relations and LAMOST data.}             
\label{all_temp}      
\begin{center} 
\scriptsize
\begin{tabular}{c c c c c c c}    
\hline\hline     
     J0158     &        J0732 & J1013        & J1324        & J1524        & Ref \\
\hline
 5384(133)  & 5733(151) & 4925(167) & 5187(108) & 5336(145) & 1   \\
 5500(125)  & 5500(125) & 5000(125) & 5000(125) & 5250(125) & 2   \\
 5492(15)   & 5494(184) & 4926(140) & 4997(58)  & 5198(21)  & 3   \\
 5459(61)   & 5576(90)  & 4950(84)  & 5061(58)  & 5261(64)  & 4   \\

\hline                  
\end{tabular}
\end{center} 
%\begin{tablenotes}
%\scriptsize
%\item The enteries in column "Ref" refer to different methods used to determine the $T_{eff}$: [1] for the $(J-H)-T_{eff}$ relation of \cite{2007MNRAS.380.1230C}, [2] for SED fiiting, [3] for the average of the values listed in LAMOST catalogue, and [4] for the average of these values.
%\end{tablenotes}

{\raggedright The enteries in column "Ref" refer to different methods used to determine the $T_{eff}$: [1] for the $(J-H)-T_{eff}$ relation of \cite{2007MNRAS.380.1230C}, [2] for SED fiiting, [3] for the average of the values listed in LAMOST catalogue, and [4] for the average of these values. \par}
\end{table}

% Figure 04
\begin{figure}[!ht]
\includegraphics[width=\columnwidth]{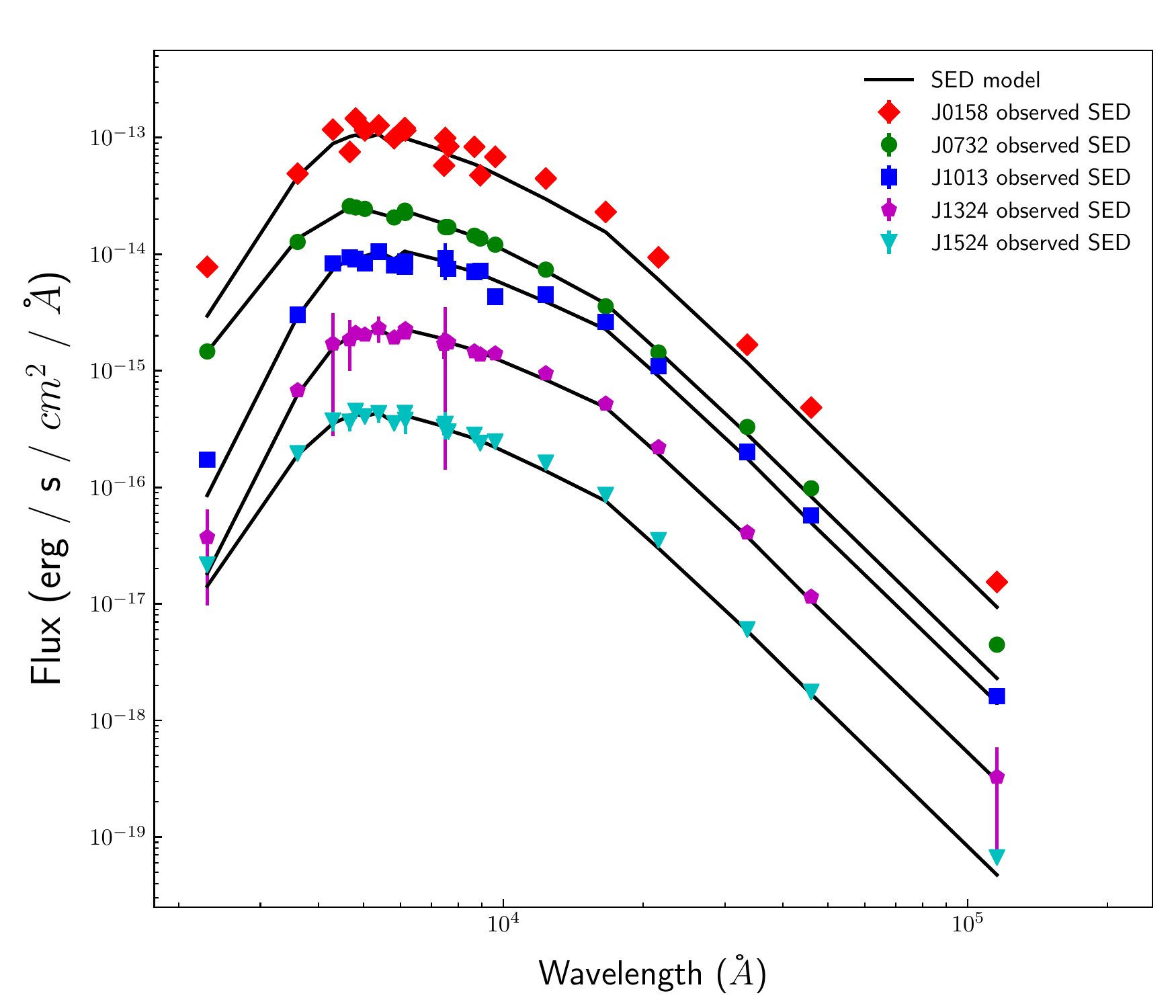}
\caption{The generated SED and fitted curve for the five systems. A vertical shift is introduced while plotting to make every SED clearly visible.
}
\label{all_sed}
\end{figure}
In the modeling process, we need to fix some of the parameters like $HJD_{o}$, period, etc. The temperature of primary and/or secondary can be fixed provided a good estimate is available. The photometric results are not affected much by small changes in the temperature but fundamental parameters like luminosity of individual components, separation between components (which we will derive from the total luminosity) and total mass of system can be affected by the change in effective temperature. To get a better estimate of the temperature, we used two different methods for the calculation of \teff\ in addition to the temperature estimates available in the LAMOST survey. First, we used the (J-H)-\teff\ relation given by \cite{2007MNRAS.380.1230C} as the J and H-band magnitudes were available for all the sources from 2MASS survey (\citealt{2006AJ....131.1163S}; [1] in Table~\ref{all_temp}). 
In the second method, we collected all the publicly available photometric data in different bands and created the spectral energy distribution (SED) for each system. The \teff\ was estimated with the help of VOSA SED fitter \footnote{http://svo2.cab.inta-csic.es/theory/vosa/} ([2] in Table~\ref{all_temp}). It is an automatic tool which gives access to different photometric catalogs, generates and fits the SED, estimates the luminosity, generates an H-R diagram, etc. The best fit model is selected via \chitwo\ minimization. We used the \logg\ and \feh\ values as listed in the LAMOST catalog. The Kurucz ODFNEW/NOVER model was used for SED fitting \citep{2008A&A...492..277B}. The fitted SEDs for all the sources are shown in Figure~\ref{all_sed}. The flux is shifted by some constant multiplication factor ($F_{J0158}\times10$, $F_{J0732}\times5$, $F_{J1013}$, $F_{J1324}\times0.2$ and $F_{J1524}\times0.04$) so that all the SEDs are clearly visible. The \teff\ values resulting from the SED fitting are 5500 ($\pm$125), 5500 ($\pm$125), 5000 ($\pm$125), 5000 ($\pm$125) and 5250 ($\pm$125) K for J0158b, J0732, J1013, J1324 and J1524, respectively ([2] in Table~\ref{all_temp}). The Table~\ref{all_temp} also includes the average \teff\ values as listed in the LAMOST catalogue (indicated with [3]). We used the non-weighted mean of the three different \teff\ estimates for each system ([4] in Table~\ref{all_temp}) as effective temperature of the primary component (\teffp) for the light curve modeling. Note that the subscripts 1 and 2 are used throughout the paper to refer to the primary and secondary components, respectively.

%--------------------------------
\subsection{Light Curve Solution}\label{q_param}
%--------------------------------
% Figure 05
\begin{figure*}%[!ht]
\begin{center}
\label{q_para}
\subfigure{\includegraphics[width=5.75cm, height=4.8cm]{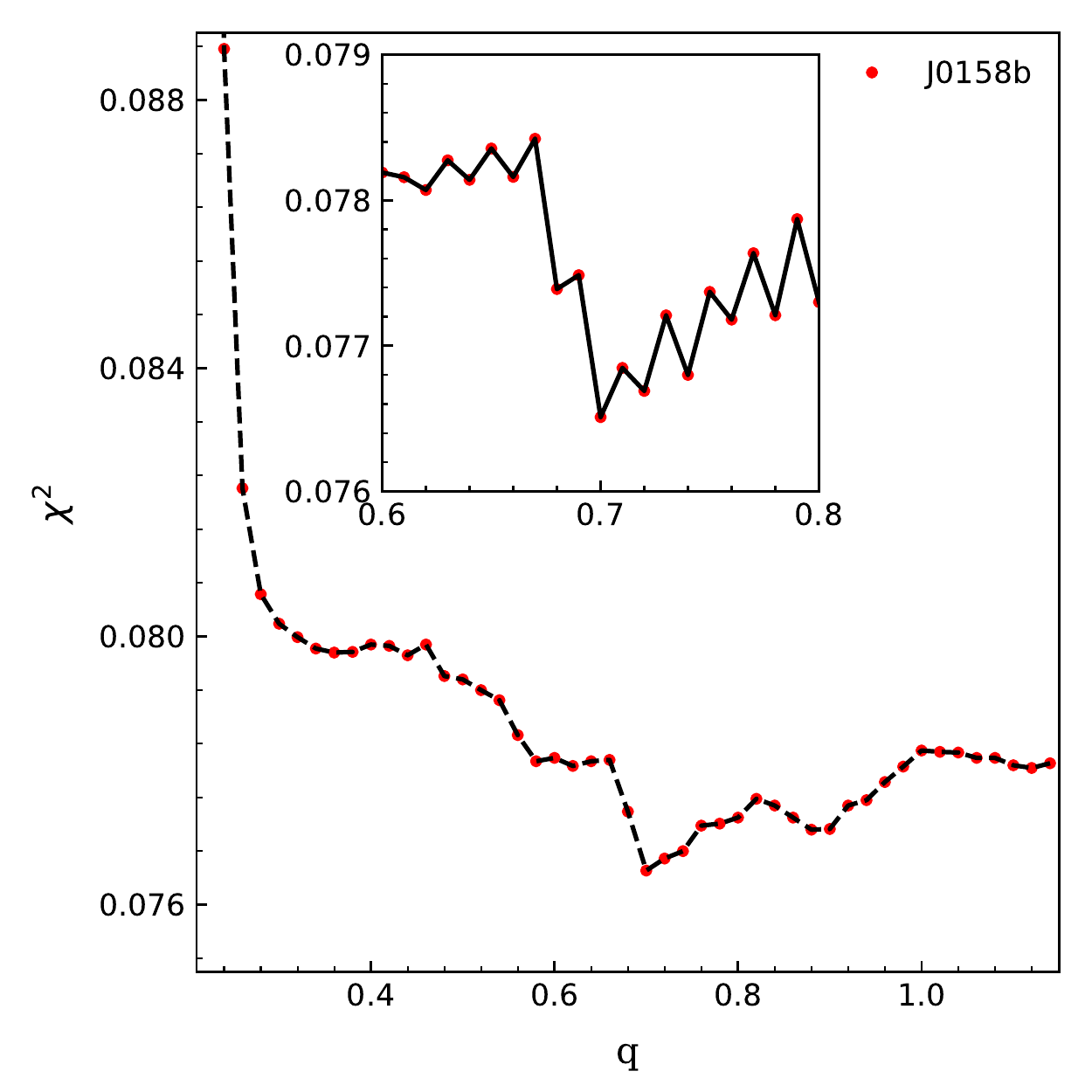}}
\subfigure{\includegraphics[width=5.75cm, height=4.8cm]{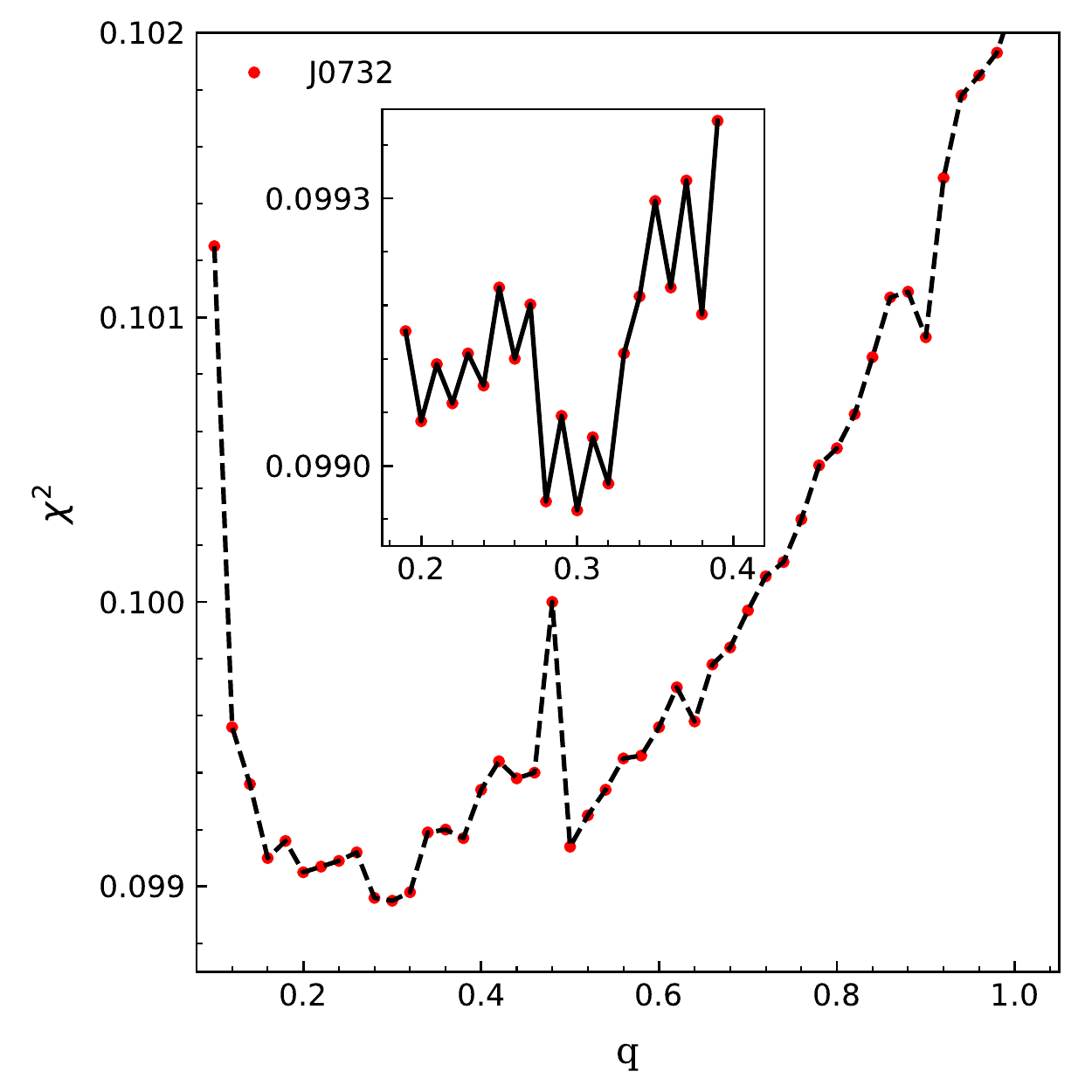}}
\subfigure{\includegraphics[width=5.75cm, height=4.8cm]{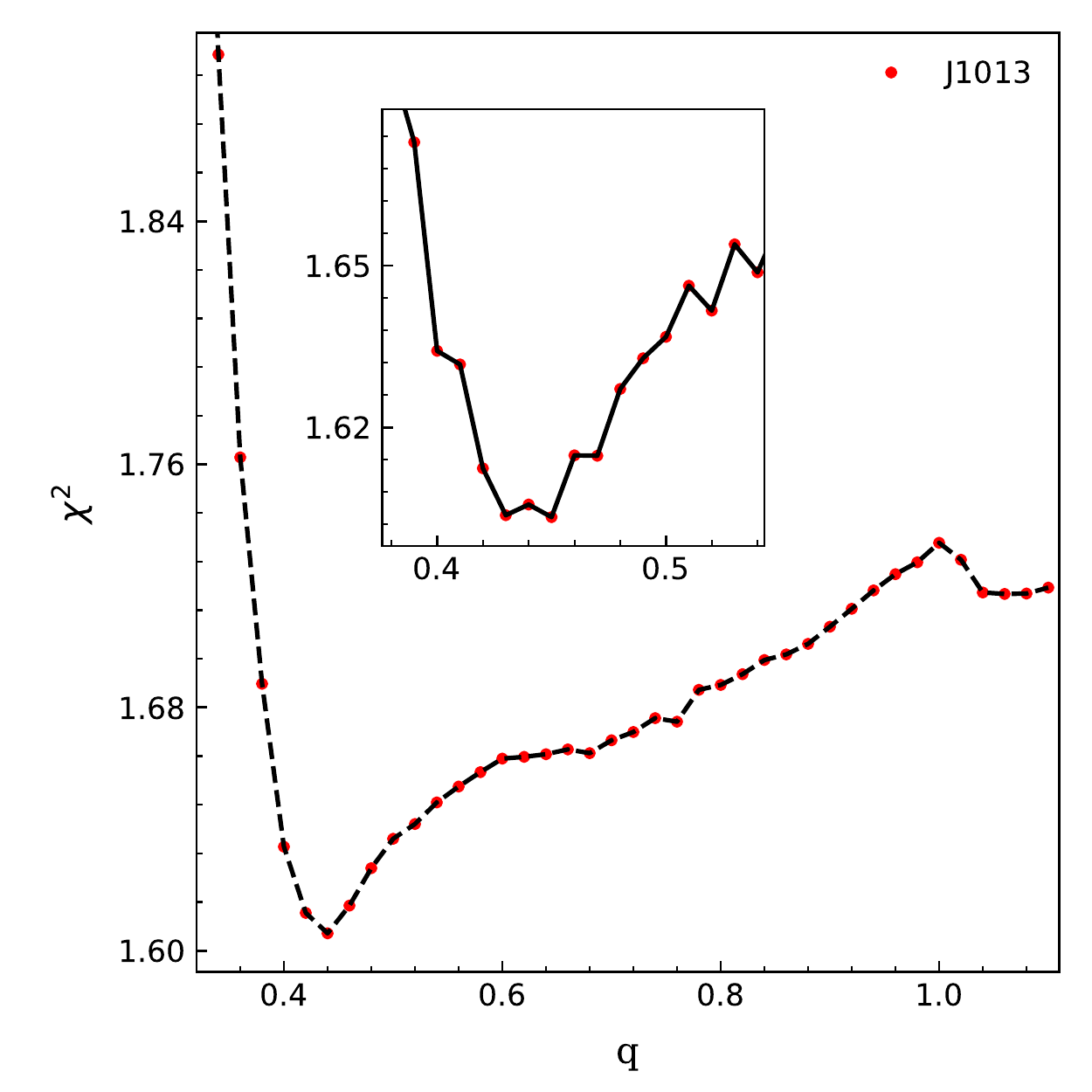}}\vspace{-0.3cm}
\subfigure{\includegraphics[width=5.75cm, height=4.8cm]{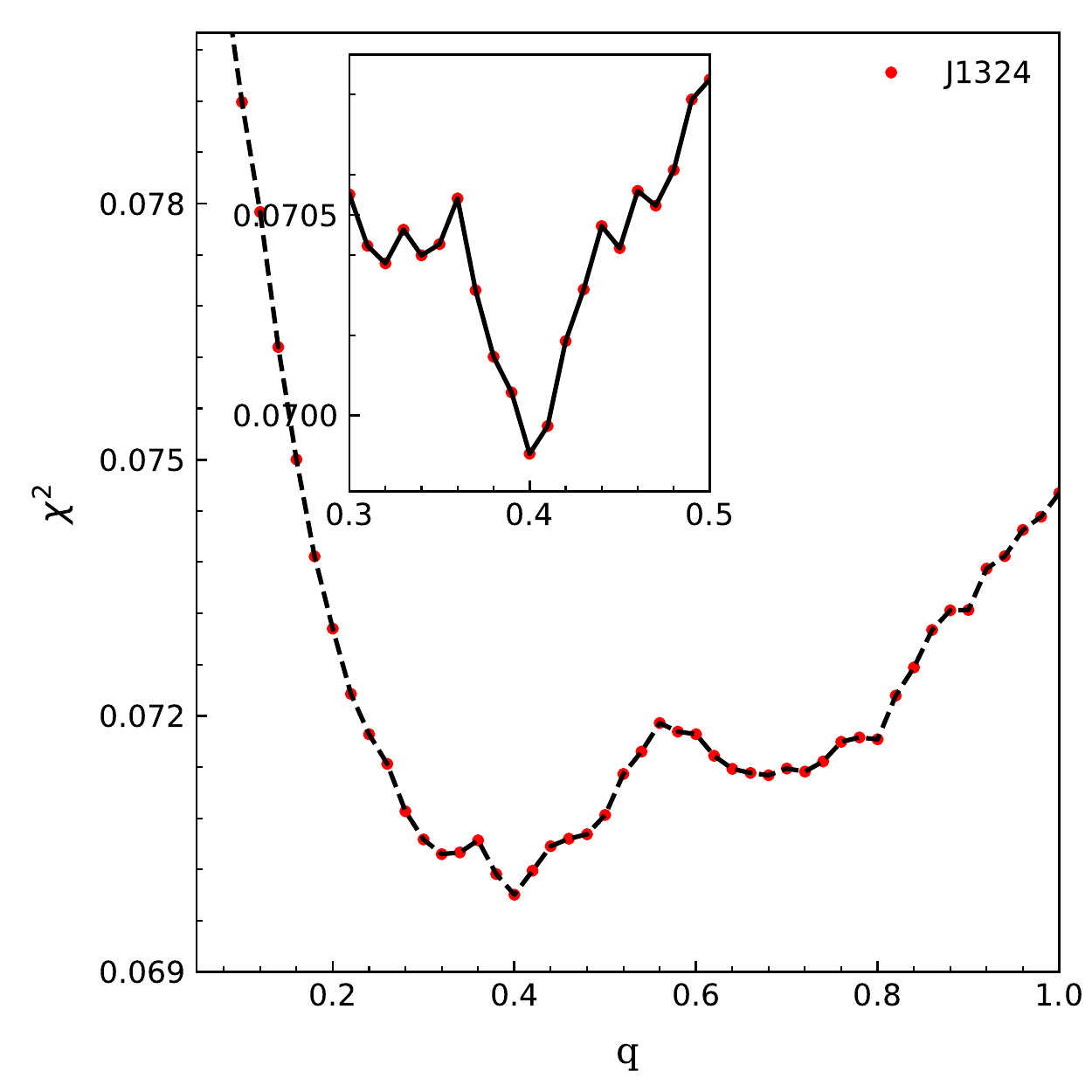}}
\subfigure{\includegraphics[width=5.75cm, height=4.8cm]{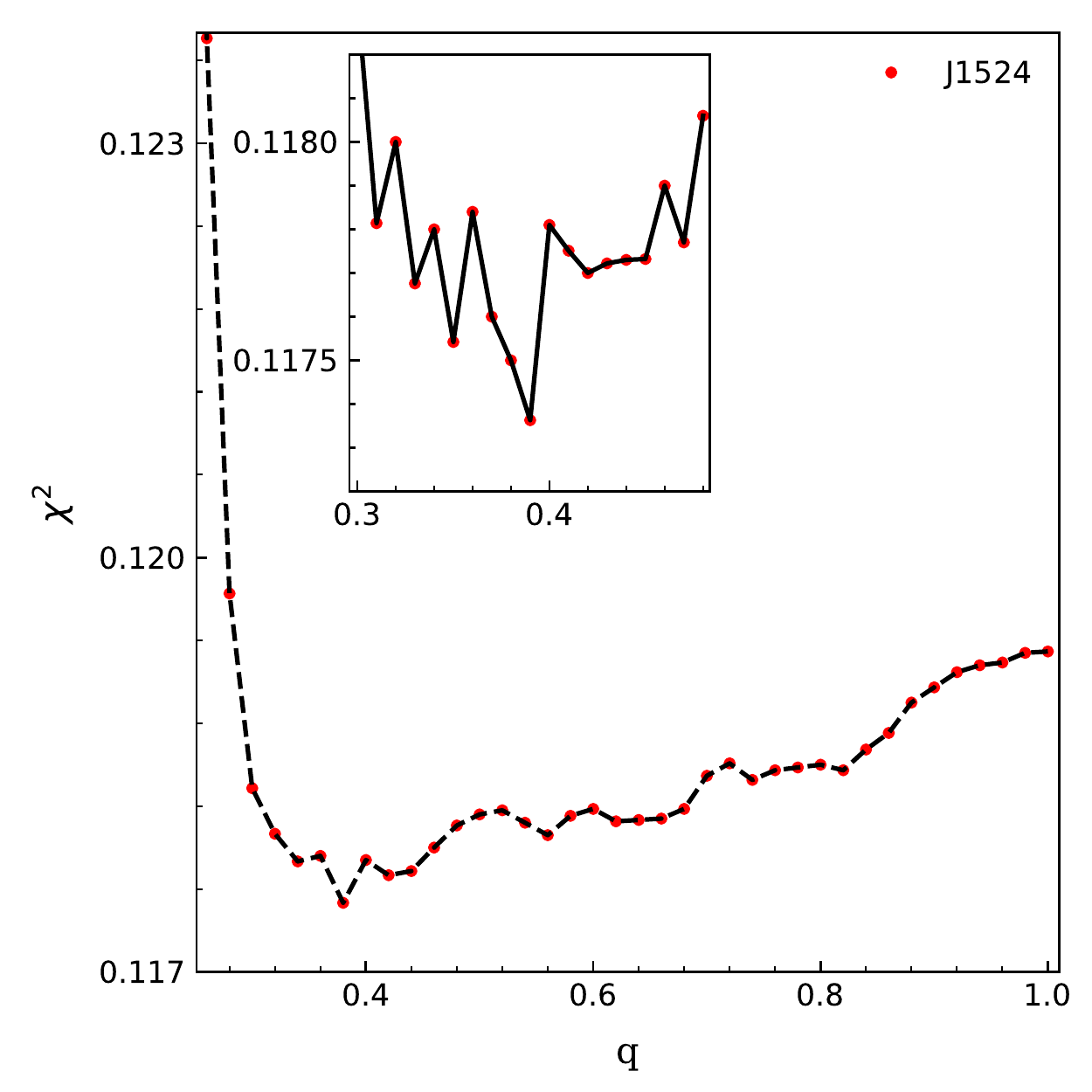}}
\caption{
The variation of \chitwo\ for different values of the mass ratio $q$ for J0158b (top left), J0732 (top middle), J1013 (top right), J1324 (bottom left), and J1524 (bottom right). A zoom of the region around the minimum \chitwo\ value with smaller increments in q (0.01) is given in the subpanels.
}
\end{center}
\end{figure*}

These sources were not observed using high-resolution spectrographs in the past, so no information was available about their RV variations. As we have been carrying out the first detailed photometric analysis of these target stars, no earlier estimates of the mass ratio ($q=M_{2}/M_{1}$) were available in the literature. We derived $q$ by applying the $q$-search technique on the available multi-band photometric LCs \citep[e.g.,][]{2016RAA....16...63J, 2017RAA....17..115J}. For the q-search, only DFOT and ST data were used. All the multiband LCs from DFOT and ST were used at the same time for q-search. The orbital period, TOMs, \teffp, gravity darkening coefficients ($g_{1}=g_{2}=0.32$), and bolometric albedos ($A_{1}=A_{2}=0.5$) were fixed while the effective temperature of the secondary component (\teffs), surface potentials ($\Omega_{1}$=$\Omega_{2}$), luminosity of the primary component ($L_{1}$), and orbital inclination ($i$)  were set free. The eccentricity ($e$), rate of orbital period change, synchronocity parameters ($F_{1}, F_{2}$) and third light ($l_{3}$) were fixed to zero during analysis. The code automatically selects and modifies the limb darkening coefficients from the \cite{1993AJ....106.2096V} tables after each iteration. While applying the $q$-search process, different models were created with mass ratios varying from 0.04 to higher values in small steps of 0.02. Following every 30 iterations, the parameter set was altered randomly by $\pm5\%$ of their actual value. For each $q$ value we used this random shift in parameters for 20 times. Therefore every 600 ($20\times30$) iterations gave the best fit model for a given $q$ input. It is expected that the residuals after subtraction of the synthetic LCs from the observed LCs would decrease as soon as the $q$ would approach the true mass ratio. The $q$ vs. \chitwo\ plot is shown in Figure~\ref{q_para}. The region around the minima is further explored with smaller step size of 0.01 as shown in the zoomed sub-panels of Figure~\ref{q_para}.
As all the targets are EWs according to the CRTS catalog, we used the contact mode during q-search. On the basis of fill-out factor it was found that the system J1324 was on the edge of contact geometry. Therefore, same procedure of q-search was repeated but this time with semi-detached mode. The semi-detached mode gave a slightly better fit as compared to the contact mode (\chitwo decreased to 0.0399 from 0.0413) for J1324. Hence for J1324 analysis, the semi-detached mode of PHOEBE was used.
 The initial estimate for $q$ are 0.70($\pm$0.02), 0.30($\pm$0.01), 0.45($\pm$0.02), 0.40($\pm$0.01), and 0.39($\pm$0.02) for J0158b, J0732, J1013, J1324, and J1524, respectively. As the more massive component was considered as primary for all systems, the resulting $q$ values were below 1.
The mass-ratio estimated using the q-search and other parameters corresponding to that mass-ratio are the initial estimates. To refine the model parameters and determined the uncertainties more robustly, Markov Chain Monte Carlo (MCMC) technique was used. A python script which wraps PHOEBE and the EMCEE code by \cite{2013PASP..125..306F} was used for this purpose \citep{2005ApJ...628..426P}. The EMCEE is a python based sampler which uses the MCMC methods that remains unaffected by affine tranformations \citep{2010CAMCS...5...65G}. We adjusted five parameters, q, i, \teffs, $\omega_{1/2}$ and $L_{1}$ during the MCMC run. The number of walkers were set to 100 and 5000 iterations were used. The parameters determined from the q-search were used as priors. A boundary limit was used for all the five parameters.  The chosen number of chains and iterations evaluated 500000 models. We discarded first 10000 iterations out of these 500000 models. The output of initial chains was used as starting point for latter iterations. The mean and the standard deviation were used as the final parameters and uncertainties. The Figures~\ref{corna} and ~\ref{cornb} at the end show the posterior distributions and the parameter correlations for five parameters. The temperature of secondary (\teffs), surface potential of primary/secondary ($\Omega_{1/2}$), inner/outer critical roche equipotential ($\Omega_{in/out}$), luminosity ratio in different bands ($L_{1}/L_{T}$), relative radii of primary/secondary ($r_{1/2}$) in units of semi-major axis, and fill-out factor ($f$) obtained from the LC modeling are given in Table~\ref{mod_para}.

\begin{figure*}[!ht]
\begin{center}
\label{scan}
\subfigure{\includegraphics[width=16cm, height=3cm]{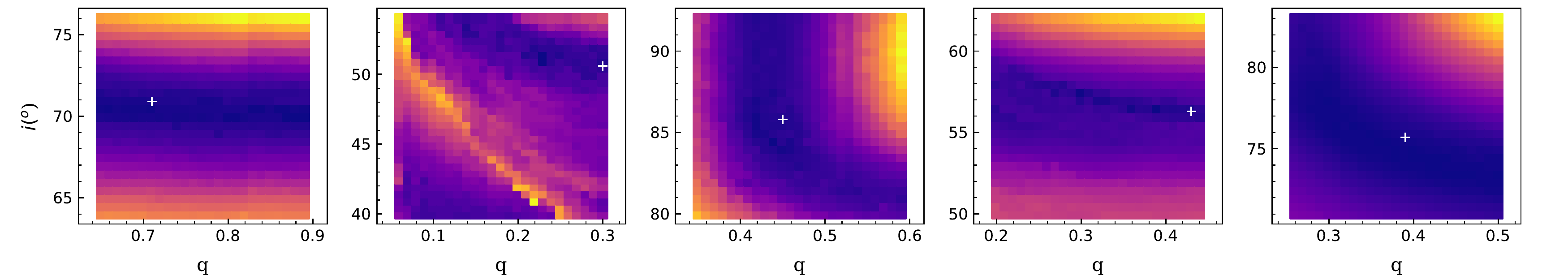}}
\subfigure{\includegraphics[width=16cm, height=3cm]{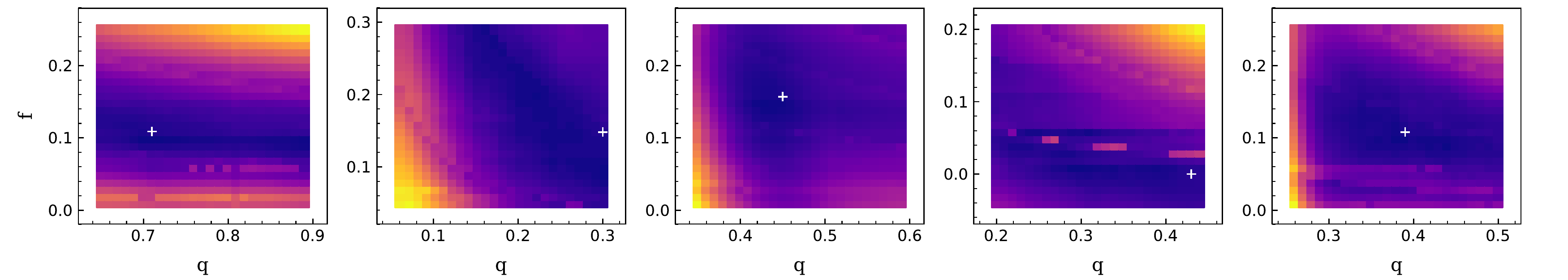}}\vspace{-0.0cm}
\subfigure{\includegraphics[width=16cm, height=3cm]{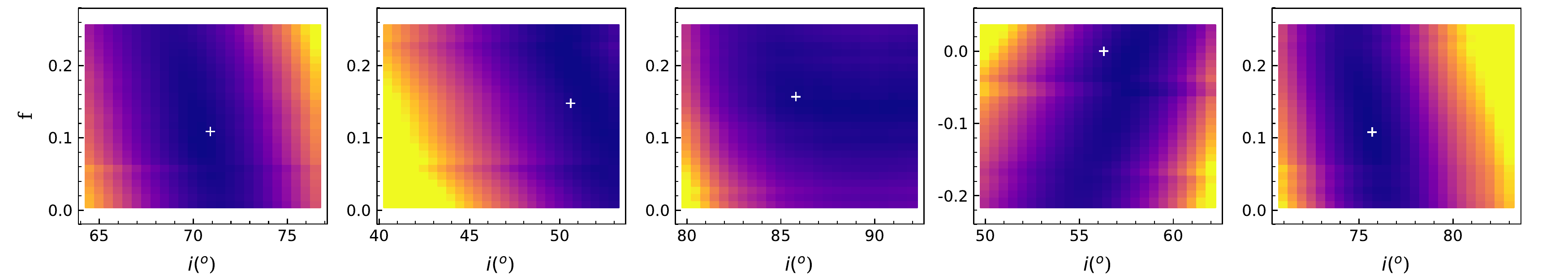}}
\caption{ 
The $q$-$i$, $q$-$f$ and $i$-$f$ parameter space around the adopted solutions (shown by white '+' sign) for all the targets. 
The color yellow is used for the highest \chitwo\ values and blue for the lowest ones. The panels in column 1 to 5 correspond to J0158b, J0732, J1013, J1324 and J1524, respectively.
}
\end{center}
\end{figure*}

The parameter space around the final solutions was also analysed. The region around final parameters was divided into a grid and multiple iterations were used to find the minimum $\chi2$ for each point on this parameter grid \citep{2005ApJ...628..426P}. Although different minimization techniques can give best solution for appropriately fed datasets after multiple iterations but sometimes minimizer can stuck in local minima also. As mentioned by \cite{2005ApJ...628..426P}, the global minima can have lots of local minima within itself as it is very flat. The parameter degeneracy and noise on the datapoints can make it difficult for the  minimization techniques to converge the solution to the global minimum. Therefore, random $\pm5\%$ kicks in the parameter values were applied after every 40 iterations with the help of PHOEBE-scripter. These random kicks help to knock the solution parameter set out of local minima (if present) and facilitate the converging process to reach the global minima. The process helps to increase the convergence efficiency of the minimization method.
%%%
 The 2-d parameter space around the solutions in different combinations of ($q$-$i$), ($q$-$f$) and ($i$-$f$) are shown in Figure~\ref{scan}. Figure~\ref{scan} shows the variation of \chitwo\ in the $q$-$i$, $q$-$f$, and $i$-$f$ parameter space. The variation of color from yellow to blue in the Figure~\ref{scan} shows changing \chitwo. It is clearly visible that the adopted final solution falls in the bluer region of the parameter space and hence corresponds global minima region.
%
% Figure 06
%\begin{figure*}[!ht]
%\begin{center}
%\label{uniq}
%\subfigure{\includegraphics[width=6cm, height=5cm]{fig06a.pdf}}
%\subfigure{\includegraphics[width=6cm, height=5cm]{fig06b.pdf}}\vspace{-0.3cm}
%\caption{
%The sample histograms showing the process of parameter and associated error determination using heuristic scanning and parameter kicking for the mass ratio $q$ (left) and the inclination $i$ (right) of J0158b.
%}
%\end{center}
%\end{figure*}
%
% Figure 6
%
%Figure 7
\begin{figure*}[!ht]
\begin{center}
\includegraphics[width=16cm, height=4.6cm]{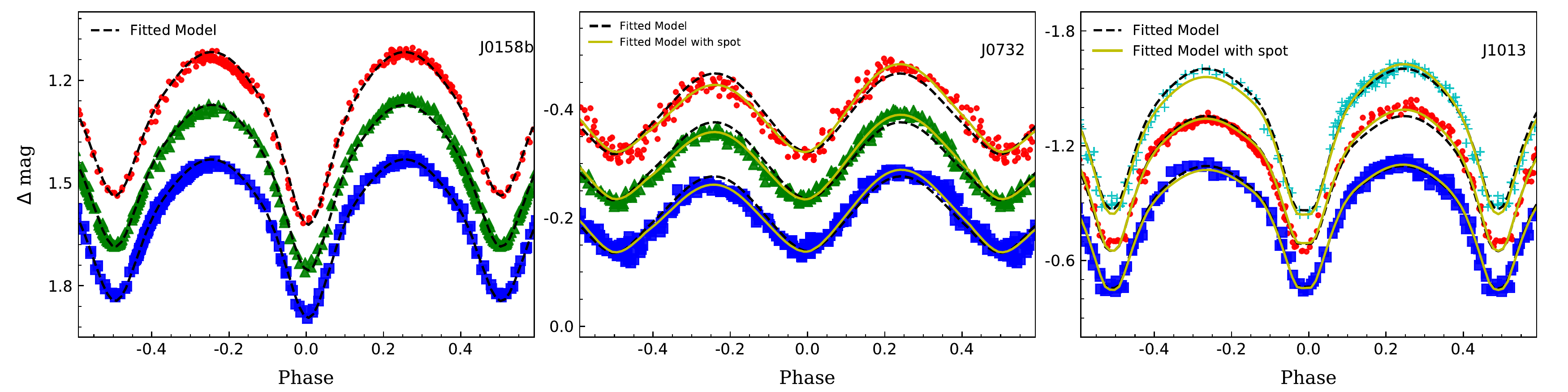}
\includegraphics[width=16cm, height=4.6cm]{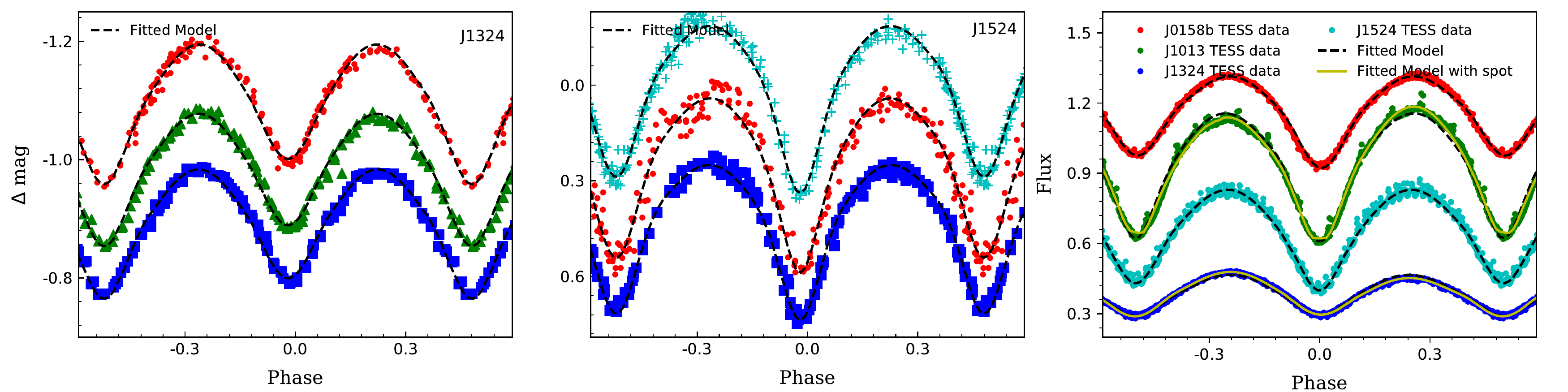}
\caption{The synthetic LCs (black dashed lines for model without spot and yellow continuous for model with spot) are over-plotted to the observed data in different bands as shown by cyan + marker (B), red dots (V), filled green triangles (R) and filled blue squares (I) dots. The last plot shows TESS observations corresponding model generated synthetic and different colors indicates different targets.
}
\label{mo_fit}
\end{center}
\end{figure*}

For J0158b, the components temperature ratio (\teffs/\teffp) $\,$  is $\sim$ 0.95. The secondary temperature is almost 266 K less than the primary temperature. The secondary minima are almost 0.07 mag dimmer than the primary minima. The primary component is bigger as well as hotter in this system, so, it is essentially a A-subtype EW. As the model "Overcontact not in thermal contact" is used for the photometric LC analysis, the potential $\Omega$ of both the primary and secondary star are kept the same ($\Omega_{1}=\Omega_{2}$). They are found to be 3.22. The resulting fill-out factor $f$ of $10.9\%$ indicates that it is a shallow contact type system. The same $q$ is used for the modeling of the TESS LC of J0158b. We used the same procedure and found a slightly different inclination $i$. The best fit model on TESS data have an orbital inclination and \teffs\ of 69.24$^{\circ}$ and 5074\,K, respectively.

For the system J0732, the amplitude of variation is 0.16 mag which is smaller than other targets. The mass ratio $q$ is determined to be 0.296($\pm$0.005) and the orbital inclination $i$ is estimated to be 50.6($\pm$0.2)$^{\circ}$. Unlike J0158b, the secondary component in J0732 has almost similar temperature to the primary component. The components temperature ratio in this case is found to be 1.002. The fill-out factor $f$ for this system is determined as 9.4$\%$. The difference between the two maxima (around orbital phase 0.25 and -0.25) is visible in Figure~\ref{mo_fit}. This difference is almost 0.03 mag.  The brightness is expected to be equal around phase -0.25 and 0.25 because the same amount of cross-section is visible in both the cases. This asymmetrical behavior could originate from a spot on the surface of one of the components. In most cases, a time series of high-resolution spectra is required to map the surface of each component. This process is known as Doppler Imaging. It helps to determine the position, size, distribution, and motion of the spot. However, it is possible to determine accurate spot parameters solely based on photometric data with an accuracy of 0.1\,mmag or better \cite{1999TJPh...23..357E}. As the other system parameters could be affected by the use of incorrect spot parameters, they were fixed while determining the spot features through \chitwo\ minimization. We tried different positions for the cool/hot spot on primary/secondary and on the basis of minimum \chitwo, spot was placed at co-latitude $45^{\circ}$ and longitude $90^{\circ}$ of the primary. The temperature ratio and size were determined to be 0.83 and $15^{\circ}$, respectively.

The system J1013 is a totally eclipsing binary system with an inclination of 85.76($\pm$0.09)$^{\circ}$. Both the components have almost similar temperature: the components temperature ratio is calculated as 1.01. The fill-out factor $f$ is determined as 15.7$\%$. This system also shows a difference in brightness at orbital phases -0.25 and 0.25. The system is $\sim$0.07 mag brighter at orbital phase 0.25. We tried a cool spot on secondary as well as a hot spot on the primary component but the hot spot on primary gave a better fit. Therefore, a hot spot was used on primary and placed at co-latitude $86^{o}$ and longitude $80^{o}$. The temperature ratio and size is determined to be 1.12 and $17^{o}$, respectively. In case of the TESS data analysis, $i$ and \teffs\ are determined as $86^{\circ}$ and 5003\,K, respectively. The luminosity ratio ($L_{1}/L_{T}$) in the TESS band is estimated as 0.651. The position of hot spot is slightly different as determined using the TESS LC. The hot spot is estimated to be at co-latitude $84^{o}$ and longitude $71^{o}$, in case of TESS data. The system shows W-subtype characteristics as the bigger component is cooler. 

The system J1324 also has a low orbital inclination of 56.3($\pm$0.2)$^{\circ}$. The temperature of the secondary is higher than the primary component by $\sim$ 7 $\%$ i.e. nearly 334\,K. The J1324 is expected to be a semi-detached system. The bigger component i.e. the primary of the system is cooler than the secondary. Therefore, the system is classified as W-subtype. As only a marginal asymmetry level is observed in the DFOT data ($\sim$0.008\,mag), no spot is used in the LC analysis. In the case of the TESS data, the asymmetry is larger. We therefore included a spot on the primary in the modeling of the TESS LC, reducing the \chitwo\ from 0.076 to 0.023. The co-latitude, longitude, radius and temperature ratio are determined as  $57^{\circ}$, $279^{\circ}$, $15^{\circ}$ and 0.88, respectively. While using the TESS LC, the components temperature ratio is determined as 1.03. The \teffs\ estimated by TESS LC is $\sim$ 250\,K  below the \teffs\ value determined from the DFOT LCs.

\begin{figure*}[!ht]
\begin{center}
\includegraphics[width=17.0cm, height=8.5cm]{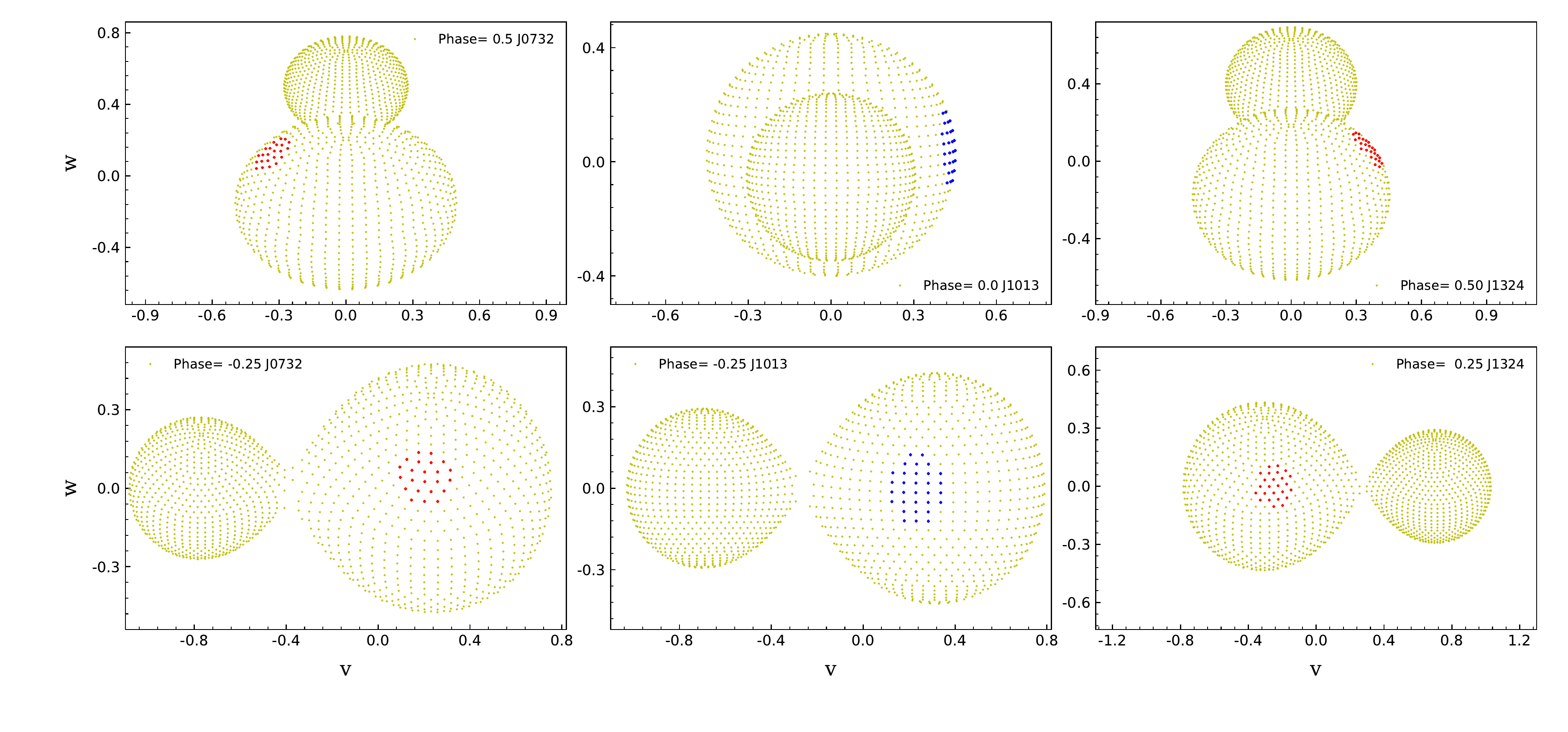}
\caption{The spot distribution on surface of J0732 (left), J1013 (middle) and J1324 (right) on two different orbital phases (top and bottom). Both the hot and cool spots are shown with the help of blue and red region, respectively.
}
\label{spots}
\end{center}
\end{figure*}

For J1524, $q$ and $i$ are found to be 0.389 and $\sim76^{\circ}$, respectively. The primary is hotter than the secondary by 109\,K. The J1524 system is a A-subtype system. The fill-out factor $f$ is 10.8\%. The secondary temperature is calculated as 5152\,K using the TESS LC. With the TESS data, we find the orbital inclination as $\sim74^{\circ}$. The parameters determined using the DFOT and ST photometric LCs are given in Table~\ref{mod_para}. The radii given in Table~\ref{mod_para} are relative radii of the components in the unit of semi-major axis (A). They are determined by $(r_{pole}\times r_{side}\times r_{back})^{1/3}$.  The synthetic LCs along with observed LCs are shown in Figure~\ref{mo_fit}. The spot distributions on the surface of J0732, J1013 and J1324 are shown in Figure~\ref{spots}.

% Figure 08
%

%Table 06
\begin{table}[!ht]
\caption{The light curve solution for targets derived using DFOT and ST observations. For J1324 the fill-out factor of the secondary is given. The errors are in parentheses. }             
\label{mod_para}      
\centering
\scriptsize
%\begin{tabular}{l c c c c c}    
\begin{tabular}{p{.45in}p{.48in}p{.48in}p{0.48in}p{0.48in}p{0.48in}}
\hline\hline     
Parameters               & J0158b     & J0732      & J1013      & J1324      & J1524      \\
\hline
q                        & 0.71(1)    & 0.296(15)  & 0.45(1)    & 0.43(1)    & 0.39(1)    \\
i ($^{\circ}$)           & 70.9(1)    & 50.6(2)    & 85.8(1)    & 57.4(1)    & 75.7(3)    \\
\teffs/\teffp            & 0.95(1)    & 1.002(17)  & 1.01(2)    & 1.07(1)    & 0.98(1)    \\
$\Omega_{1}$             & 3.22(1)    & 2.44(1)    & 2.73(1)    & 2.74       & 2.63(1)    \\
$\Omega_{2}$             & 3.22(1)    & 2.44(1)    & 2.73(1)    & 2.72(2)    & 2.63(1)    \\
%$\Omega_{in}$            & 3.26443    & 2.457374   & 2.772504   & 2.746670   & 2.655576   \\
%$\Omega_{out}$           & 2.85686    & 2.272638   & 2.502582   & 2.483872   & 2.417735   \\
$L_{1}/L_{T}$(B)         & -----      & -----      & 0.633      & -----      & 0.728      \\
$L_{1}/L_{T}$(V)         & 0.643      & 0.755      & 0.658      & 0.596864   & 0.729      \\
$L_{1}/L_{T}$(R)         & 0.630      & 0.753      & -----      & 0.610890   & -----      \\
$L_{1}/L_{T}$(I)         & 0.614      & 0.745      & 0.659      & 0.611605   & 0.713      \\
$r_{1}$                  & 0.418      & 0.491      & 0.463      & 0.453383   & 0.471      \\
$r_{2}$                  & 0.361      & 0.289      & 0.321      & 0.309162   & 0.308      \\
$f (\%)$                 & 10.9(2.5)  & 09.4(5.4)  & 15.7(3.7)  & 8.4(7.6)   & 10.8(4.2)  \\
\hline                  
\end{tabular}
\vspace{1ex}
\end{table}

%section 05
%============================
\section{Physical Parameters}\label{phy_para}
%============================

The physical parameters, semi-major axis (A), mass of each component ($M_{1}, M_{2}$), radius of components ($R_{1}, R_{2}$) and luminosity of components ($L_{1}, L_{2}$) were determined using the photometric solutions and GAIA parallax as explained by \cite{2019RAA....19...14K}. The adopted steps are given below:

1.) The peak apparent magnitude ($m_{v}$) was determined by gaussian fitting around phase -0.25/0.25 of the CRTS V-band LC. With the help of GAIA parallax ($\pi$), extinction ($A_{V}$) and the BC (different for each system due to different temperature, surface gravity and metallicity), apparent magnitudes were converted to the bolometric magnitudes ($M_{bol}$).

2.) The total luminosity ($L_{T}=L_{1}+L_{2}$) of the system was determined from $M_{bol}$. As the luminosity ratio was already available from LC modeling, it was used to determine the luminosity of individual components.

3.) For determining A, we assumed general luminosity-temperature-radius relation i.e. $L_{i} \propto T_{i}^4 \times R_{i}^2$. The relative radii of each component ($r_{i}$) was determined by PHOEBE in units of A. With the help of $L_{i}$, $r_{i}$ and $T_{i}$, A was determined. The radius of each component was determined by $r_{i}\times A$ 

4.) The Kepler's third law was used to find the total mass ($M_{T}$) of the system by using the period (P) and A. The primary and secondary masses were determined using $q$. The detailed procedure and equations involved in the parameter determination can be seen in \cite{2020Ap&SS.365...71L} and \cite{2021AJ....161..221P}. The physical parameters for all the systems are given in Table~\ref{abs_para}. The values in parenthesis are the associated errors in the last digits.
%Figure 09
\begin{figure*}[!ht]
\begin{center}
\includegraphics[width=18cm, height=5.5cm]{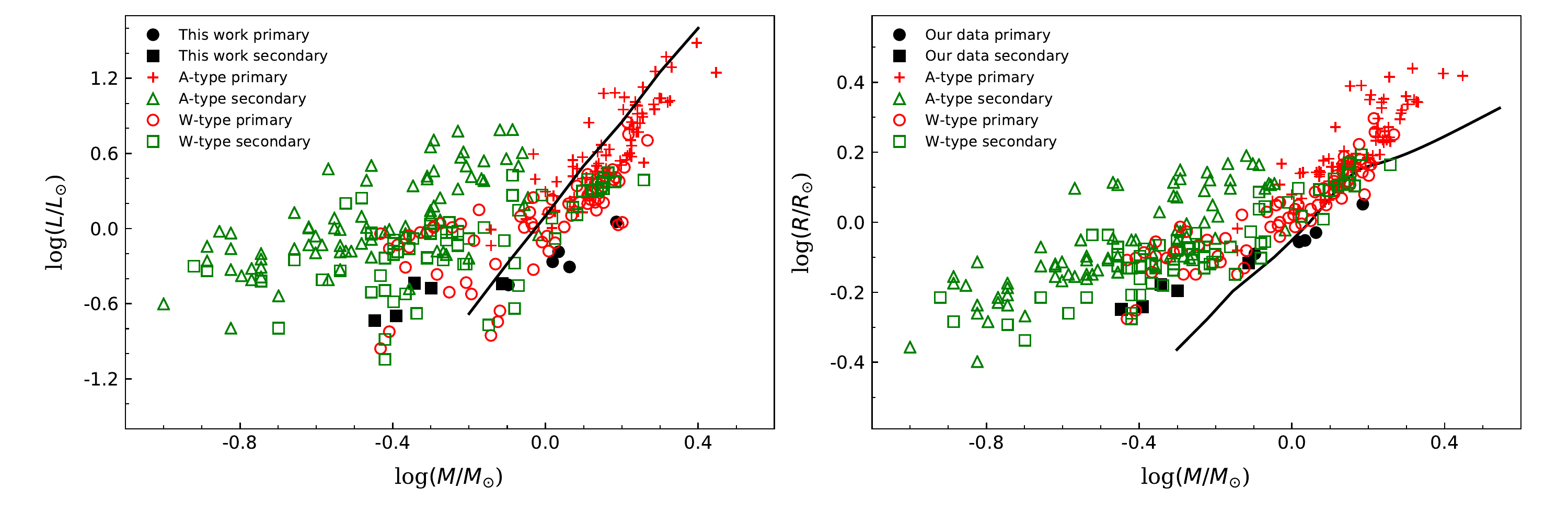}
\caption{ M-L and M-R relation for W- and A-subtype EWs. The black continuous line is ZAMS for $Z$\,=\,0.014. The systems studied in present work are shown in filled black circles (primary components) and squares (secondary components).}
\label{evol}
\end{center}
\end{figure*}
%Table 07
\begin{table}[!ht]
\caption{The absolute parameters for the five systems studied in the present work.}             
\label{abs_para}
\scriptsize
\centering          
%\begin{tabular}{l c c c c c}
\begin{tabular}{p{.50in}p{.44in}p{.44in}p{0.44in}p{0.44in}p{0.44in}}    
\hline\hline     
Parameters               & J0158b   & J0732    & J1013    & J1324    & J1524     \\
\hline
a ($R_{\odot}$)          &2.12(4)   &2.29(8)   &1.75(4)   &2.06(4)   &1.86(5)    \\
$M_{1}$ ($M_{\odot}$)    &1.08(6)   &1.5(2)    &0.81(6)   &1.16(7)   &1.04(6)    \\
$M_{2}$ ($M_{\odot}$)    &0.77(4)   &0.45(5)   &0.36(3)   &0.51(3)   &0.41(2)    \\
$R_{1}$ ($R_{\odot}$)    &0.89(2)   &1.13(4)   &0.81(2)   &0.93(2)   &0.88(2)    \\
$R_{2}$ ($R_{\odot}$)    &0.76(1)   &0.66(2)   &0.56(1)   &0.64(1)   &0.57(1)    \\
$L_{1}$ ($L_{\odot}$)    &0.65(1)   &1.13(6)   &0.354(4)  &0.49(1)   &0.54(1)    \\
$L_{2}$ ($L_{\odot}$)    &0.36(1)   &0.37(2)   &0.184(2)  &0.33(1)   &0.201(3)   \\
\hline                  
\end{tabular}
\vspace{1ex}

\end{table}

We collected the parameters of EWs which were derived using photometric and spectroscopic observations from the literature \citep[e.g.,][]{2011MNRAS.412.1787D, 2013MNRAS.430.2029Y}. These well-characterized systems as well as present results are plotted in $M$-$L$ and $M$-$R$ diagrams. It can be seen in Figure~\ref{evol} that the primary components of W and A-subtype EWs are close to ZAMS line. The secondaries of W-subtype are more distant to the ZAMS line as compared to the secondaries of A-subtype systems. The structure and evolution of secondaries is different from the primaries. The secondaries of J0732, J1013, J1324 and J1524 are more luminous for their main sequence (MS) equivalents. Both the components of J0158b are close to the ZAMS. The present targets follows the similar trend as followed by other well studied EWs.

%Section 06
%===========================
\section{Mass Transfer Rate}\label{ma_tr}
%===========================

The period analysis performed in Section~\ref{orpe} shows that the systems J1013 and J1524 exhibit change in \porb. Although there are different processes which can affect the orbital period, mass transfer between the components is the prime reason in most of the cases. The effect of processes like Gravitational wave radiation (GWR) on the orbital period is usually very small. For J1013 the possible orbital period change due to GWR is estimated to be -2.631$\times10^{-16}$\,days/year while the observed rate of change of period is -2.552$\times10^{-7}$\,days/year. Similarly, the orbital period changes that can happen due to magnetic braking is found to be -4.035$\times10^{-8}$\,days/year. The orbital period change due to the magnetic braking is almost 16\% of the observed change in J1013. The expected orbital period change due to GMR and magnetic braking is estimated as -3.533$\times10^{-16}$\,days/year and -5.130$\times10^{-8}$\,days/year, respectively, in case of J1524. Hence, for J1524, the orbital period change due to the angular momentum loss via magnetic braking is almost 76\% of the observed orbital period change rate. Therefore, in this system, magnetic braking as well as mass transfer between components can be responsible for the detected orbital period change. With the help of mass transfer equation given by \cite{1958BAN....14..131K}, we determined the yearly rate of change of mass of primary for J1013 and J1524. It was found that a -2.199$\times10^{-7}$\,M$_{\odot}$/year rate of change of mass of primary component of system J1013 can explain the observed orbital period change. Similarly, a rate of change of mass of primary component of -6.151$\times10^{-8}$\,M$_{\odot}$/year, in case of J1524, was determined to account for the observed orbital period change. Hence, the mass may be transferring from primary to the secondary component in both the components.

% Figure 10
\begin{figure*}[!ht]
\begin{center}
\subfigure{\includegraphics[width=3.5cm,height=5cm]{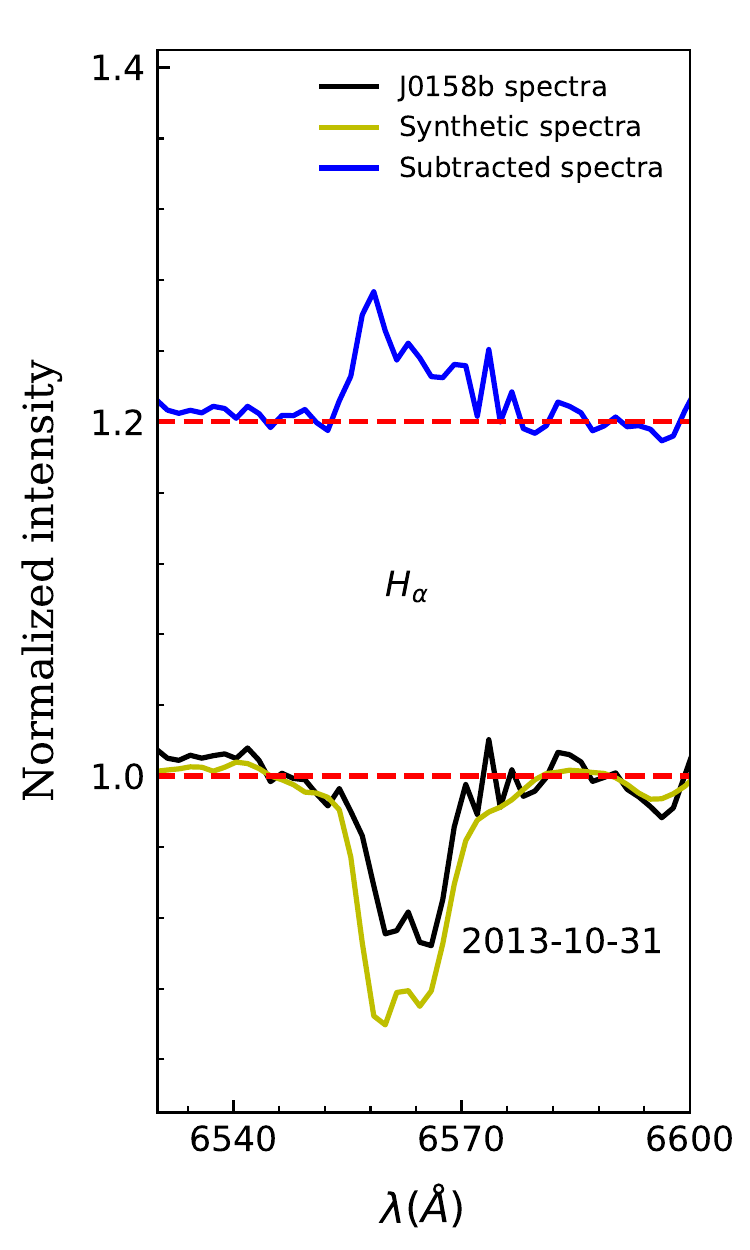}}
\subfigure{\includegraphics[width=3.5cm,height=5cm]{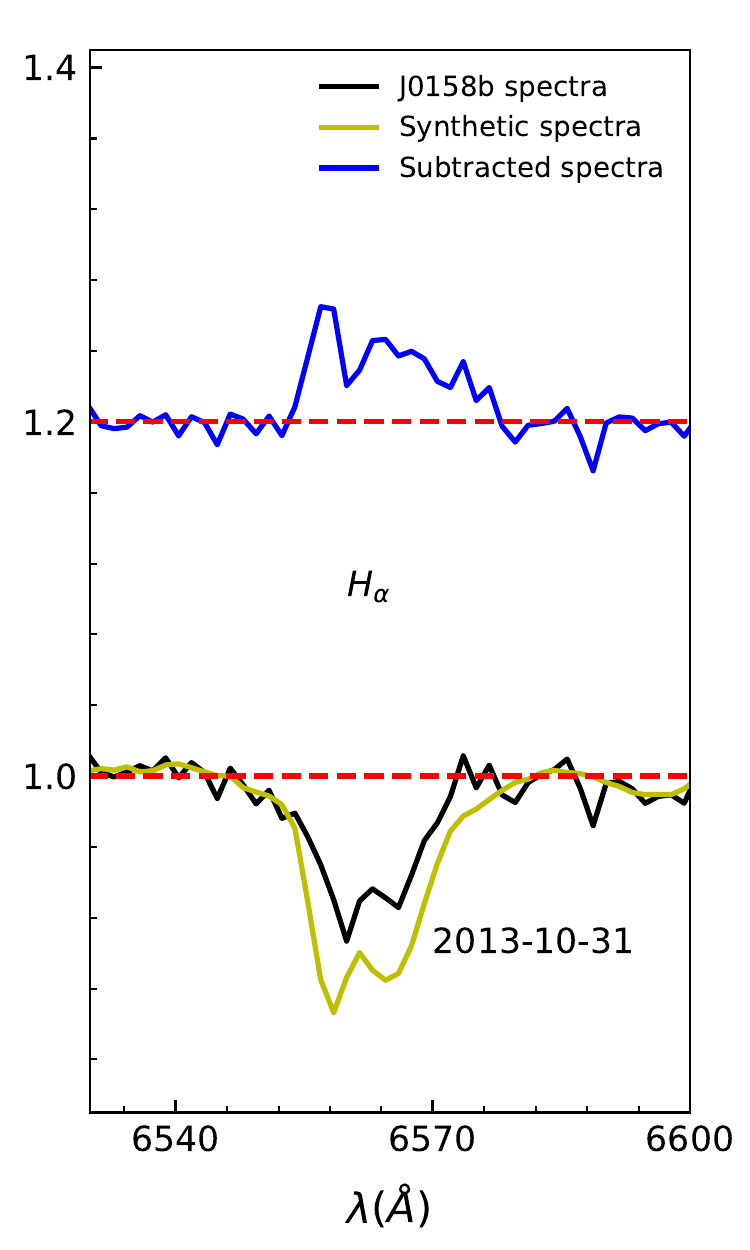}}
\subfigure{\includegraphics[width=3.5cm,height=5cm]{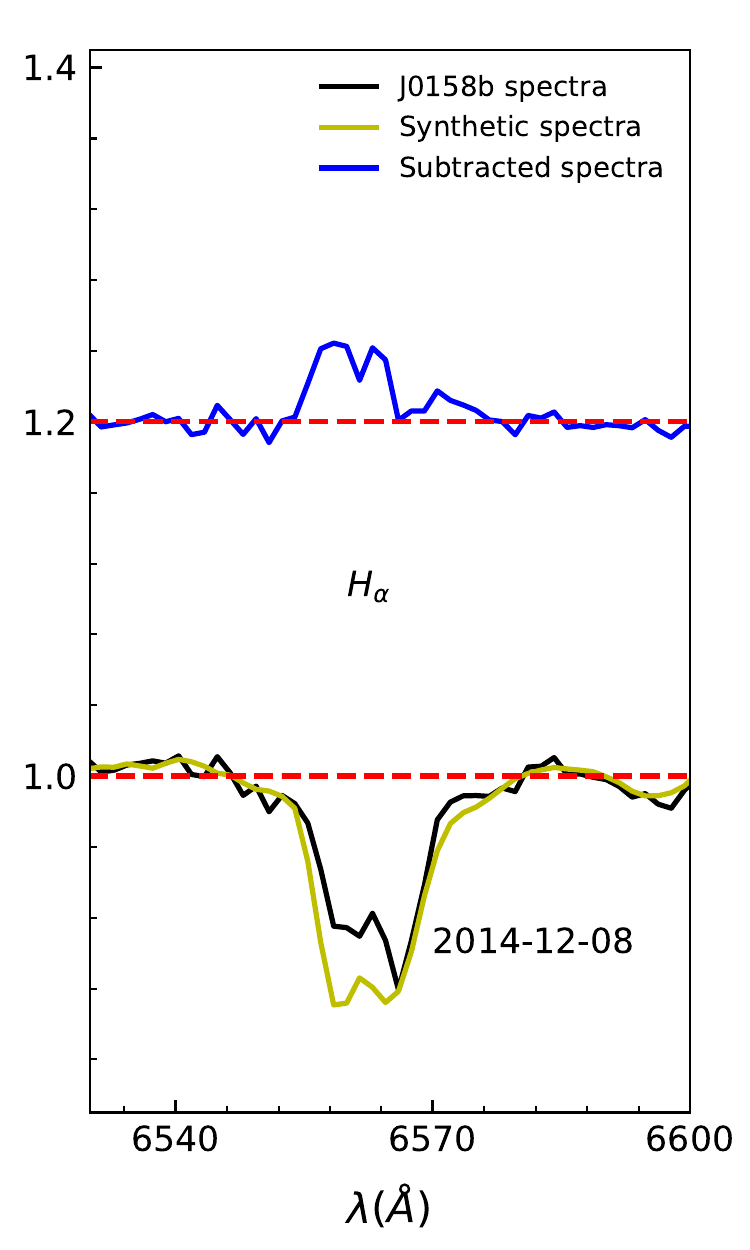}}
\subfigure{\includegraphics[width=3.5cm,height=5cm]{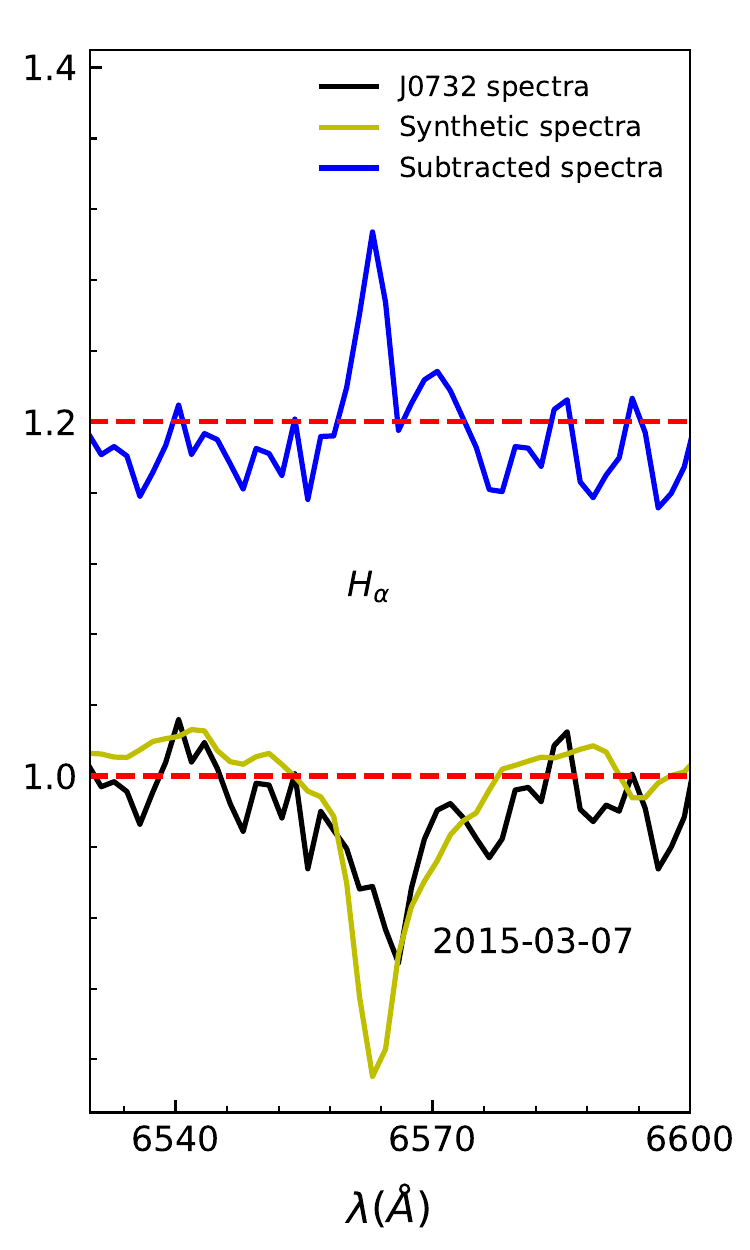}}
\caption{The \halpha\ region of low resolution LAMOST spectra (black line) and synthetic spectra (yellow line) for J0158b and J0732 are shown. The subtracted spectra in the same region is shown by a continuous blue line in each plot.}
\end{center}
\label{spec1}
\end{figure*}
%
% Figure 11
\begin{figure*}[!ht]
\begin{center}
\label{spec2}
\subfigure{\includegraphics[width=3.5cm,height=5cm]{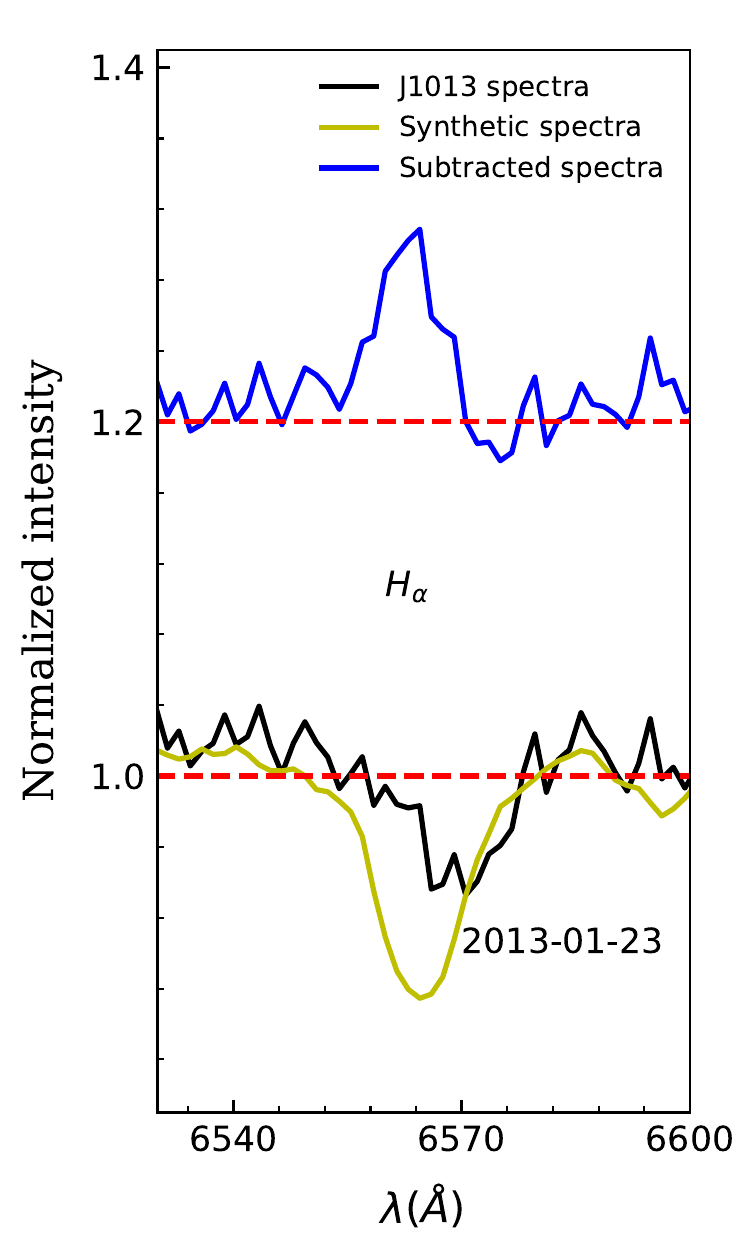}}
\subfigure{\includegraphics[width=3.5cm,height=5cm]{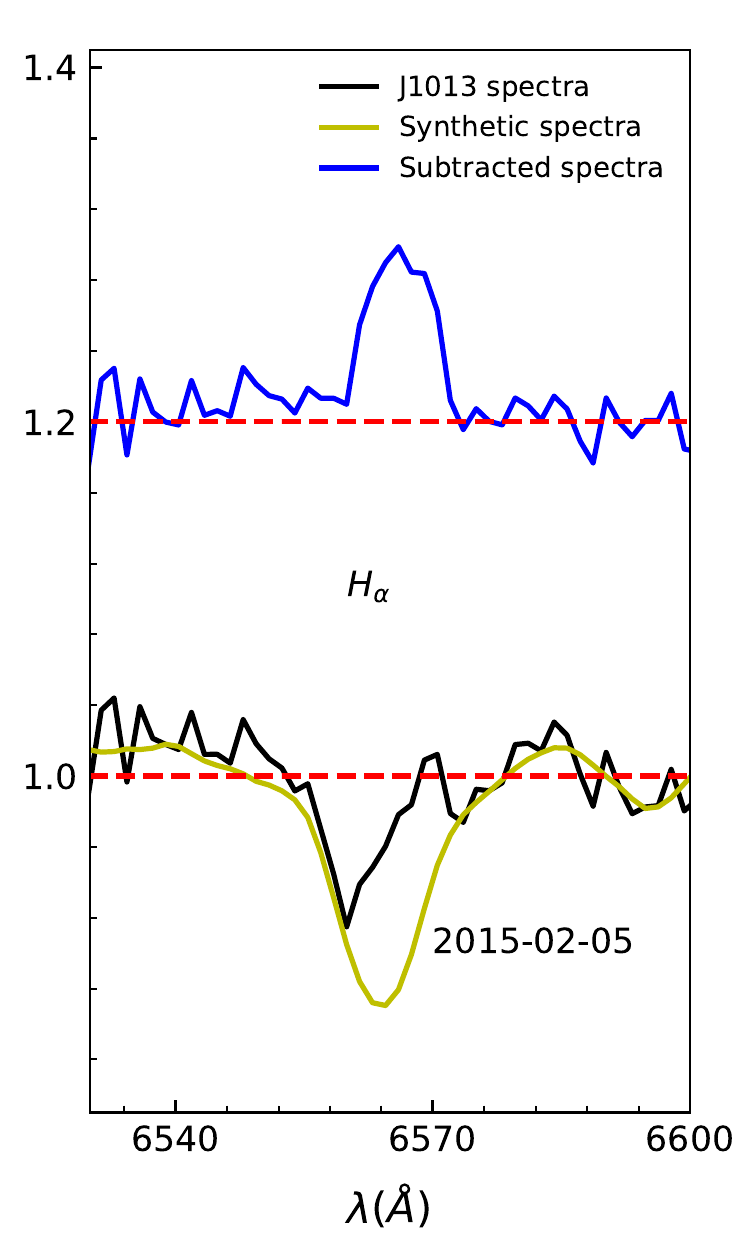}}
\subfigure{\includegraphics[width=3.5cm,height=5cm]{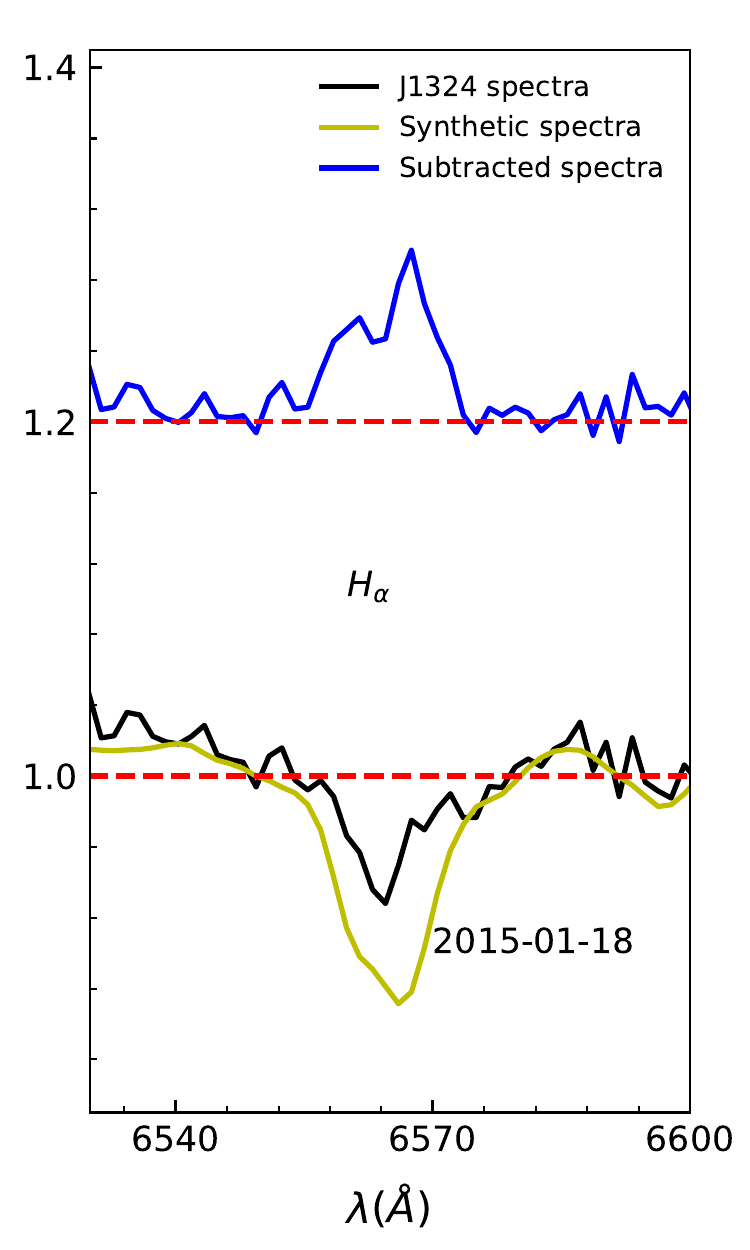}}
\subfigure{\includegraphics[width=3.5cm,height=5cm]{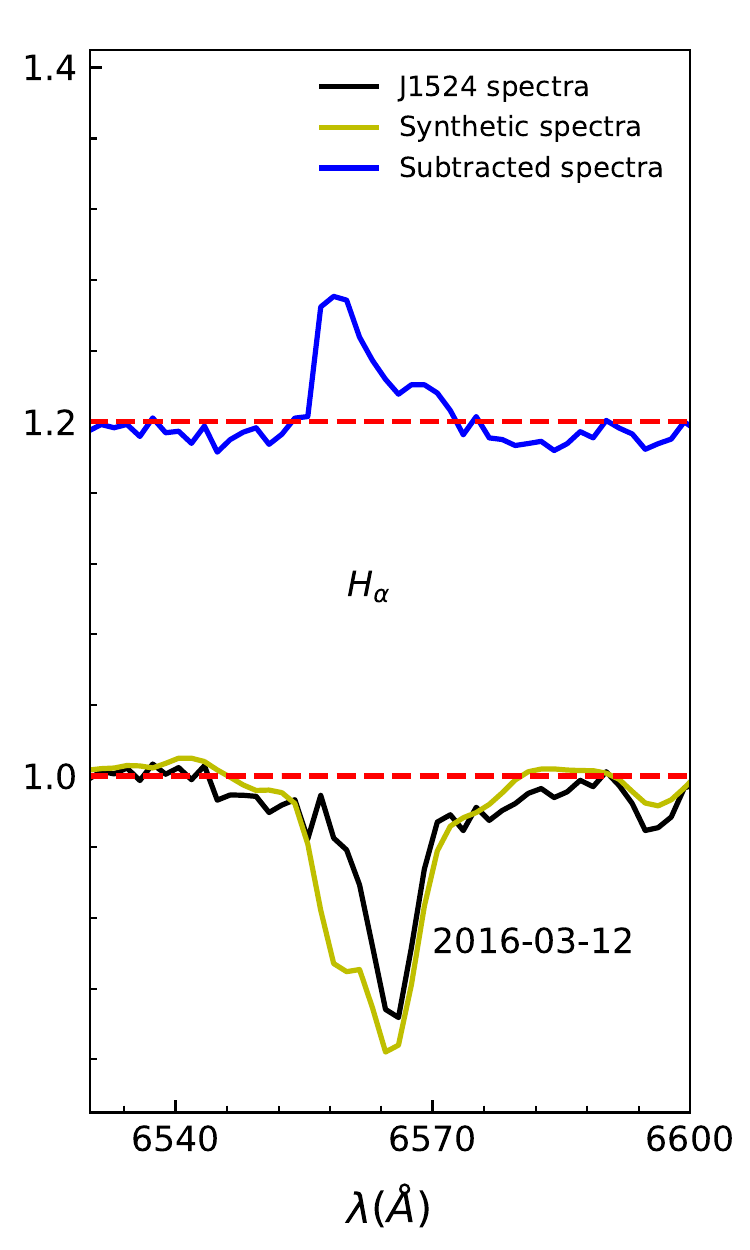}}
\caption{Same as in Figure~\ref{spec1} but for J1013, J1324 and J1524.}
\end{center}
\end{figure*}

%section 07
%=================================
\section{Chromospheric Activities}\label{ch_ac}
%=================================

The magnetic activities, as shown by the Sun (sun spots, solar flares, plages, coronal mass ejection, etc.) are also observed in late type stars. These events are due to the stellar magnetic dynamo mechanism which results from the differential rotation in the interior of the star. Phenomena like tidal forces, spinning and orbital synchronicity, and magnetic field interactions make things complex in the case of close interacting binaries. Close binary systems like RS\,CVn and BY\,Dra show observational evidence of chromospheric emission and magnetic field. \cite{1981ApJ...247..975V} studied II~Peg using multi-band photometric and low-resolution spectroscopic data and found that the strength of \halpha\ was correlated to the spot visibility. The active close binaries show emission in the Ca\,{\sc ii} and \halpha\ region of their spectra. The components of active binary systems show stronger emission as compared to the single stars with same rotational period \citep{1995A&AS..114..287M}. The level of activity can be inferred by emission or filled-in absorption lines in these wavelength regions. The H and K emission lines are primarily seen in K and M stars but they are not very common in F stars \citep{1980ARA&A..18..439L}. The contribution of flux due to the active-chromosphere can be determined by the removal of the photospheric flux from the total flux for a system. It is very hard to estimate the contribution of individual stars in the active-chromosphere in case of binary systems. The level of emission or filled-in absorption of one star can be altered due to its active/non-active companion \citep{1995A&AS..114..287M}. Such complications can be eliminated by the use of the spectral subtraction technique. It assumes that the photospheric and chromospheric flux contributions are  independent of each other which is applicable only to the localized active regions \citep{1984AJ.....89..549H, 1984BAAS...16..893B}.

The low-resolution spectra for all the targets and appropriate inactive comparison stars spectra (with spectral class similar to the target stars) selected from available catalogs \citep{1999ApJS..123..283M, 2000yCat..41420275S, 2004ApJS..152..251V} were
downloaded from the LAMOST website. All the spectra were normalized before analysis. A program called STARMOD was used for building a synthetic spectra from two non-active comparison stars. With the help of STARMOD, different pairs of comparison stars were tested as primary and secondary components. The best pair was selected on the basis of minimum \chitwo\ determined by STARMOD after comparing the synthetic and observed spectra. The template spectra were constructed for J0158 (G7-star HD\,237846 + G8-star BD\,+39\,2723),  J0732 (G7-star HD\,86873 + G5-star HD\,15299), J1013 (K3-star HD\,219829 + G9-star BD\,+12\,2576), J1324 (K1-star HD\,233826 + G9-star BD\,+12\,2576), and J1524 (G7-star HD\,237846 + G8-star BD\,+39\,2723). Figures~\ref{spec1} and ~\ref{spec2} show observed, synthetic and subtracted spectra for all the targets. The  Figures show only \halpha\ region of spectra as the Ca\,{\sc ii} H\,\&\,K and Ca\,{\sc ii} IRT wavelength regions of the spectra have a low SNR. Although small amount of excess emission can be seen in all the subtracted spectra but it should be noted that J0158b, J1324 and J1524 are the only sources with SNR greater than 100 as given in Table~\ref{tar_lamost}. Due to the low spectral resolution and closeness of the two components, the observed spectral features are a blend of both the components. Hence, it is impossible to detect the activity level of individual components and high resolution spectroscopic observations are required for this purpose.

%section 08
%===============================
\section{Results and Discussion}\label{discu}
%===============================

Photometric time series in different bands for five EWs were analyzed along with low-resolution spectroscopic data. The photometric time series data were collected from all previously available surveys to detect possible changes in their orbital period \porb. For three of the systems (J0158b, J0732, and J1324), we find no evidence for changes in their \porb\ in the timespan of the photometric observations (14-15 years) while for J1013 and J1524, the \porb\ changes are evident. The possible contribution to this orbital period change rate on the basis of different processes was calculated. For J1013, the most plausible mechanism was found to be mass-transfer from primary to the secondary while the expected period rate due to magnetic braking angular momentum loss amounts to $\sim16\%$ of the observed period changing rate. In the case of J1524, mass-transfer and/or magnetic braking angular momentum loss can be responsible for the changing period. On the basis of the calculated semi-major axis, mass, and radii, it was found that possible magnetic braking angular momentum loss can cause up to $78\%$ of the observed orbital period change rate.

For the LC modeling, the PHOEBE software was used and input parameters (like gravity darkening coefficients, surface albedos, and limb darkening coefficients) were chosen on the basis of temperature and convective envelope systems. The temperature can be determined on the basis of color-temperature relations but it can slightly deviate from the actual value if the color of the system is determined at different phases. Also, the temperature is a very important property as it is used to determine A, $M_{T}$, $R_{i}$ etc.. Therefore, to get a better estimate of the temperature of the primary component, different methods were applied and the average of all the results was taken as the final \teffp\ value. We used the model for an overcontact binary not in thermal contact in PHOEBE and the $q$-search technique to determine the mass ratio.
% The histograms were plotted for all the parameters after running the scripter for 2500 iterations. The final values and relative errors were determined by gaussian fitting to these histograms. The parameter hyperspace contains many local minima and they can have varying depth. With the help of parameters kicking, the efficiency of the minimizer can be increased to find the global minimum more easily.
The final parameters and errors are determined using the MCMC sampler with PHOEBE. To check the stability of the derived parameters, the parameter space around these solutions was scanned. For 625 models in ($q$-$i$), ($q$-$f$) and ($i$-$f$) parameter space, an iteration process was used and the best solution was derived. The results showed that the adopted solutions have a lower cost function value than most of the surrounding region. According to the best fit model, the temperature difference between the primary and secondary components was found to vary from 50 K to a few hundred K for all the systems. In the observed LCs, the depth of primary and secondary minima are different for each system. The massive and bigger components are fixed as primary component. On the basis of the classification by \cite{1970VA.....12..217B}, the systems J0158 and J1524 were classified as A-subtype while others are W-subtype EWs. Only J0158 was found to have mass ratio $q$ above 0.5. The fill-out factor $f$ was below 25\% for all the systems, so they are all classified as shallow contact type EWs.

The trustability of photometric mass ratios is an important subject. Although $q$-search using only photometric data is not as accurate as RV methods, it is used in almost every study where RV observations are unavailable. \cite{2005Ap&SS.296..221T} showed that the photometric mass ratio accuracy decreases as the system geometry changes from full eclipse to partial eclipse. Among 101 systems (59 total eclipsing and 42 partially eclipsing) collected by \cite{2021AJ....162...13L}, almost half of the partial eclipsing binary photometric mass ratios were found to differ from the ones determined from RV curves. The mismatch was as big as 0.4 in some cases where the orbital inclination angle was less than $70^{\circ}$. In present study, J1013 is the only system showing complete eclipses. Two of the other systems (J0158b and J1524) have $i >$\,$70^{\circ}$ while the the orbital inclination angle of J0732 and J1324 is below $60^{\circ}$. Hence, the mass ratio and other parameters should be considered as less reliable for the latter systems. 
The available TESS data was also modeled using the $q$ value determined from the DFOT photometric data. The LCs of J0732 and J1013 are asymmetric around phases -0.25 and 0.25, so we included a cool or hot spot on one of the components for LC modeling. As the DFOT LC of J1324 appears to be symmetric while the TESS LC shows clear asymmetries, we only included a spot for the modeling of the TESS data. Spot formation is quite common in contact binaries, especially for W-subtype EWs which are found to be more active than those of subtype A. The spots can form and disappear on the stellar surface from time to time and their lifetime can vary from days to years. Therefore it is plausible that no spots are seen in the DFOT observations while there appears to be one in the TESS data as the time of observation differs by approximately 9 months. For the determination of physical parameters of EWs, we used general relations like the luminosity-radius-temperature relation and Kepler's third law. As GAIA DR3 provides an extraordinarily precise parallax for these systems and the LC parameters available through LC modeling, the physical parameters of the components can be determined easily using these relations. Other methods involve the use of empirical relations derived from a small sample of well studied EWs. Therefore, such
relations are highly dependent on the sample. The masses and luminosities of the individual components of the systems that we investigated show that the secondaries of all the systems are more luminous and bigger in size than main sequence stars of similar masses. As luminosity and mass can be transferred from one component to the other, the position of the components can deviate from the ZAMS. Energy transfer is still under debate as it is not confirmed yet which part of the convective envelope is involved in this process.

The LAMOST spectra were used to search excess emission in the spectra. Only the region of \halpha\ could be used for this exercise. Some random peaks were observed in the subtracted spectra of all the sources but the low resolution of the LAMOST spectra prevents the estimation of the contribution of the individual components. The small excess emission peaks in the subtracted spectra of J0732 and J1013 could be noise features because the SNR of their LAMOST spectra is rather low (SNR\,$<$\,100). If we assume the emission featurs of the other systems are real, then J1524 and J1324 are likely to be more active than J0158b. Time series of high-resolution spectroscopic observations would allow to confirm or reject the suspected excess emission peaks observed in these systems and help in the calculation of the contribution of the individual components.

%Section 09
%=========================
\section{ACKNOWLEDGEMENTS}
%=========================

This work is supported by the Belgo-Indian Network for Astronomy and astrophysics (BINA), approved by the International Division, Department of Science and Technology (DST, Govt. of India; DST/INT/BELG/P-09/2017) and the Belgian Federal Science Policy Office (BELSPO, Govt. of Belgium; BL/33/IN12). This publication makes use of VOSA, developed under the Spanish Virtual Observatory project supported by the Spanish MINECO through grant AyA2017-84089. VOSA has been partially updated by using funding from the European Union's Horizon 2020 Research and Innovation Programme, under Grant Agreement nº 776403 (EXOPLANETS-A) . Guoshoujing Telescope (the Large Sky Area Multi-Object Fibre Spectroscopic Telescope LAMOST) is a National Major Scientific Project built by the Chinese Academy of Sciences. Funding for the project has been provided by the National Development and Reform Commission. LAMOST is operated and managed by the National Astronomical Observatories, Chinese Academy of Sciences. In this work we have also used the data from the European Space Agency (ESA) mission GAIA, processed by the GAIA Data Processing and Analysis Consortium (DPAC). This work also make use of the Two Micron All Sky Survey and SIMBAD database. 

\software{IRAF \citep{1986SPIE..627..733T, 1993ASPC...52..173T}, DAOPHOT \citep{1992ASPC...25..297S}, PERIOD04 \citep{2004IAUS..224..786L, 2005CoAst.146...53L}, PHOEBE v1 \citep{2005ApJ...628..426P}}.

%-------------------------------------------------------------------
\bibliographystyle{yahapj}
\bibliography{Bibilography}
\clearpage
%===============================
\begin{appendix}
The corner plots showing correlations and the posterior distributions for five parameters. 
%===============================
% Figure 12
\begin{figure*}[!ht]
\begin{center}
\label{corna}
\subfigure{\includegraphics[width=8.7cm,height=9.5cm]{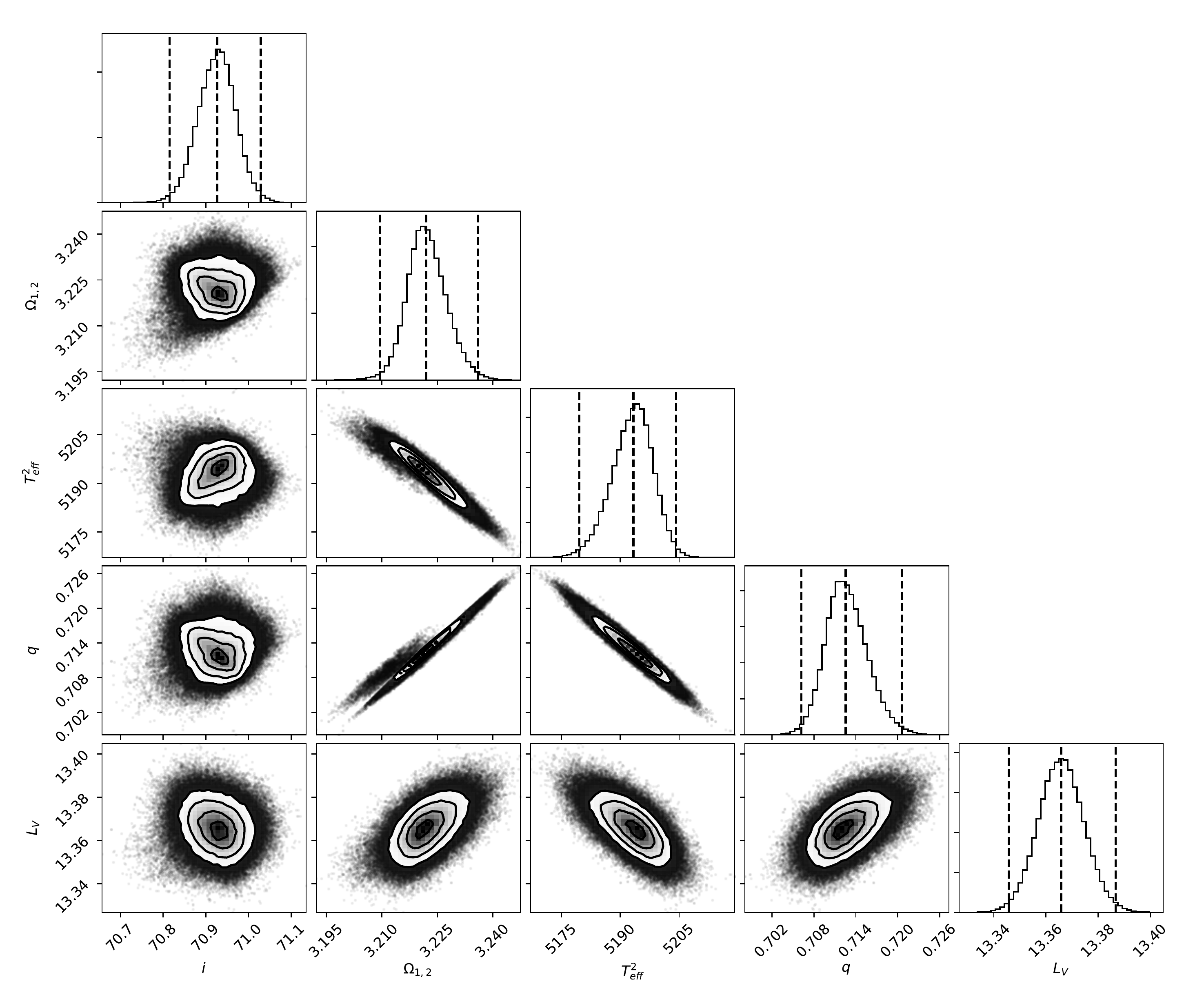}}
\subfigure{\includegraphics[width=8.7cm,height=9.5cm]{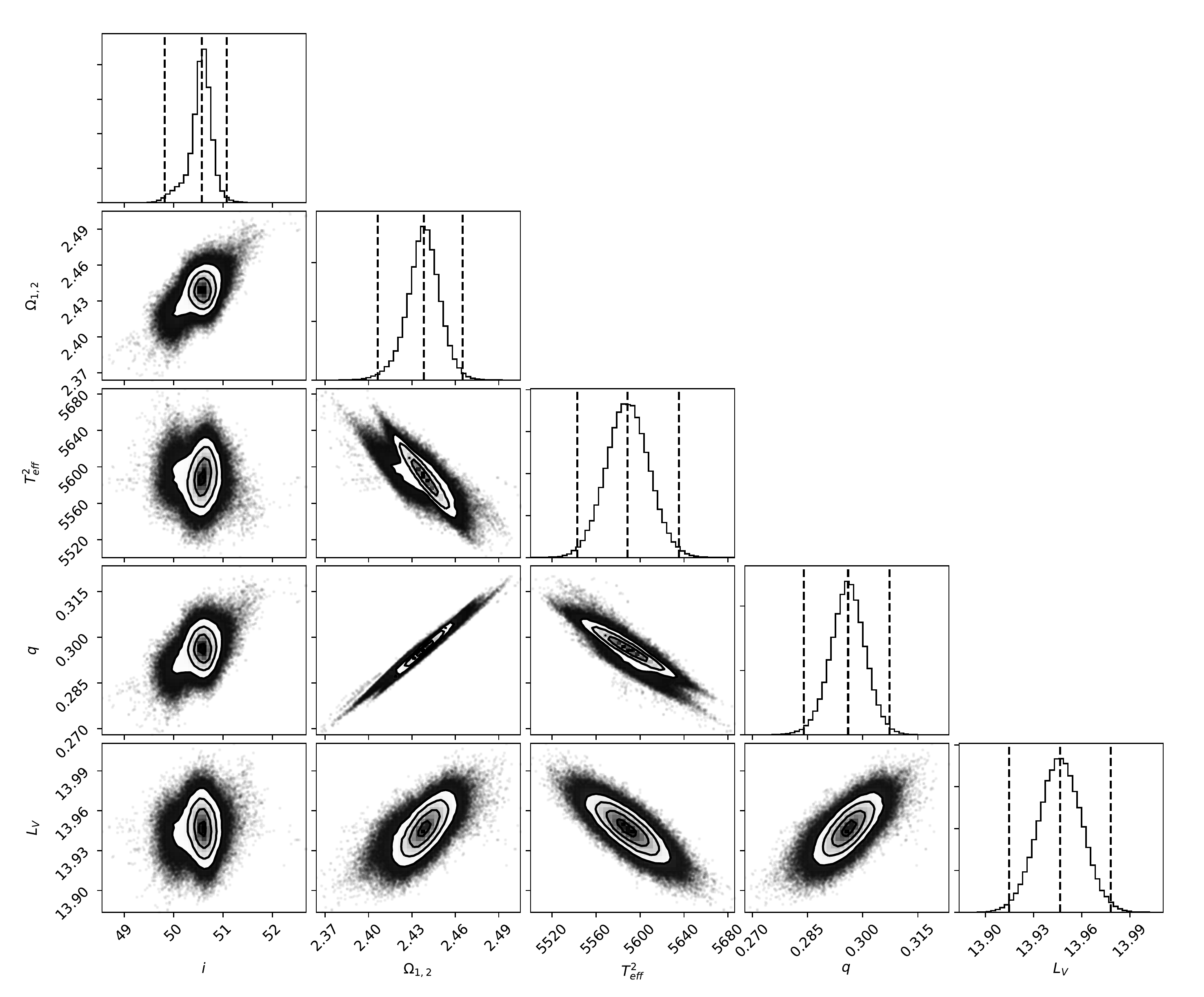}}
\caption{Corner plots for J0158 (left) and J0732 (right) showing the posterior distributions and the parameter correlations.}
\end{center}
\end{figure*}

\begin{figure*}[!ht]
\begin{center}
\label{cornb}
\subfigure{\includegraphics[width=8.7cm,height=9.5cm]{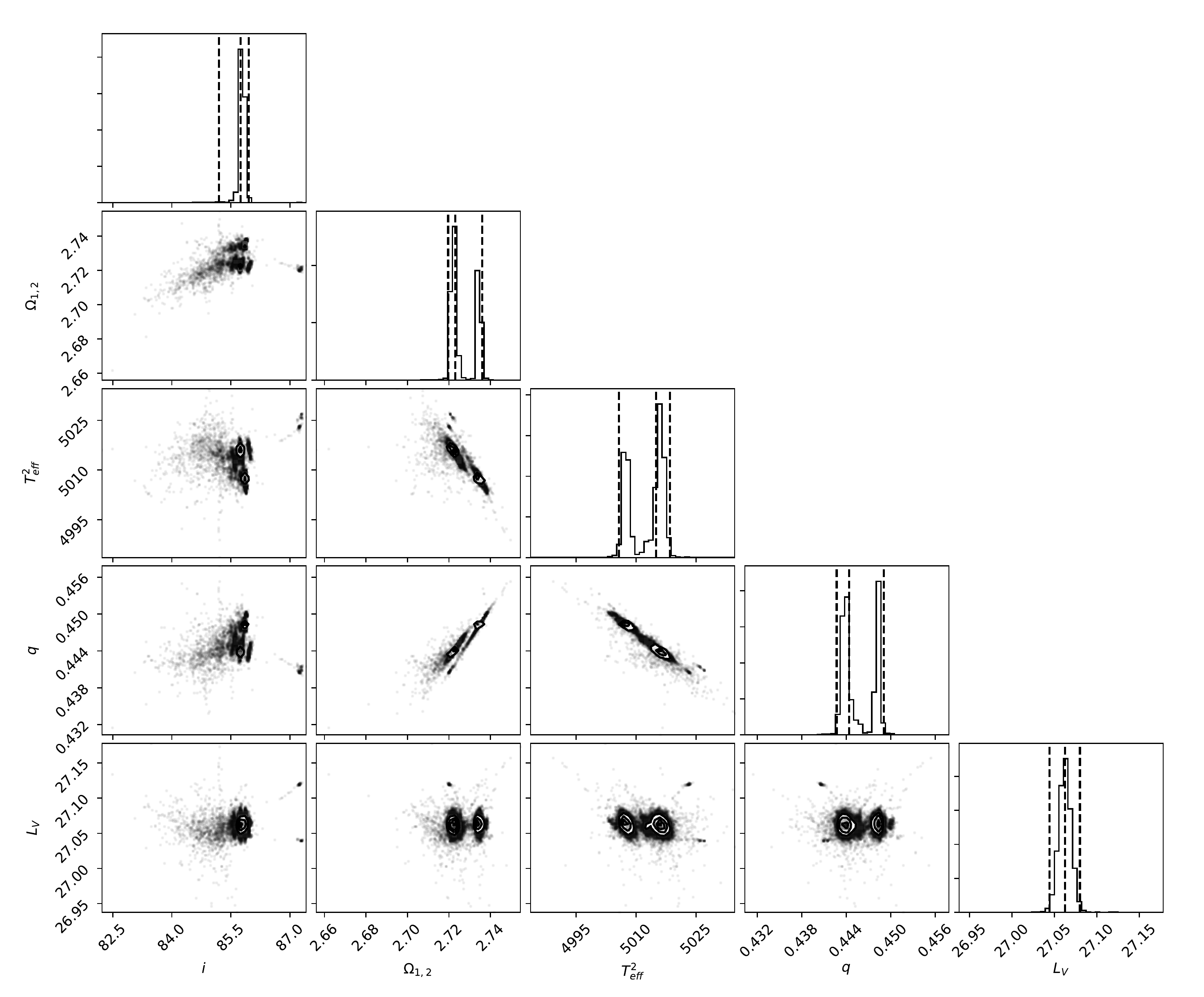}}
\subfigure{\includegraphics[width=8.7cm,height=9.5cm]{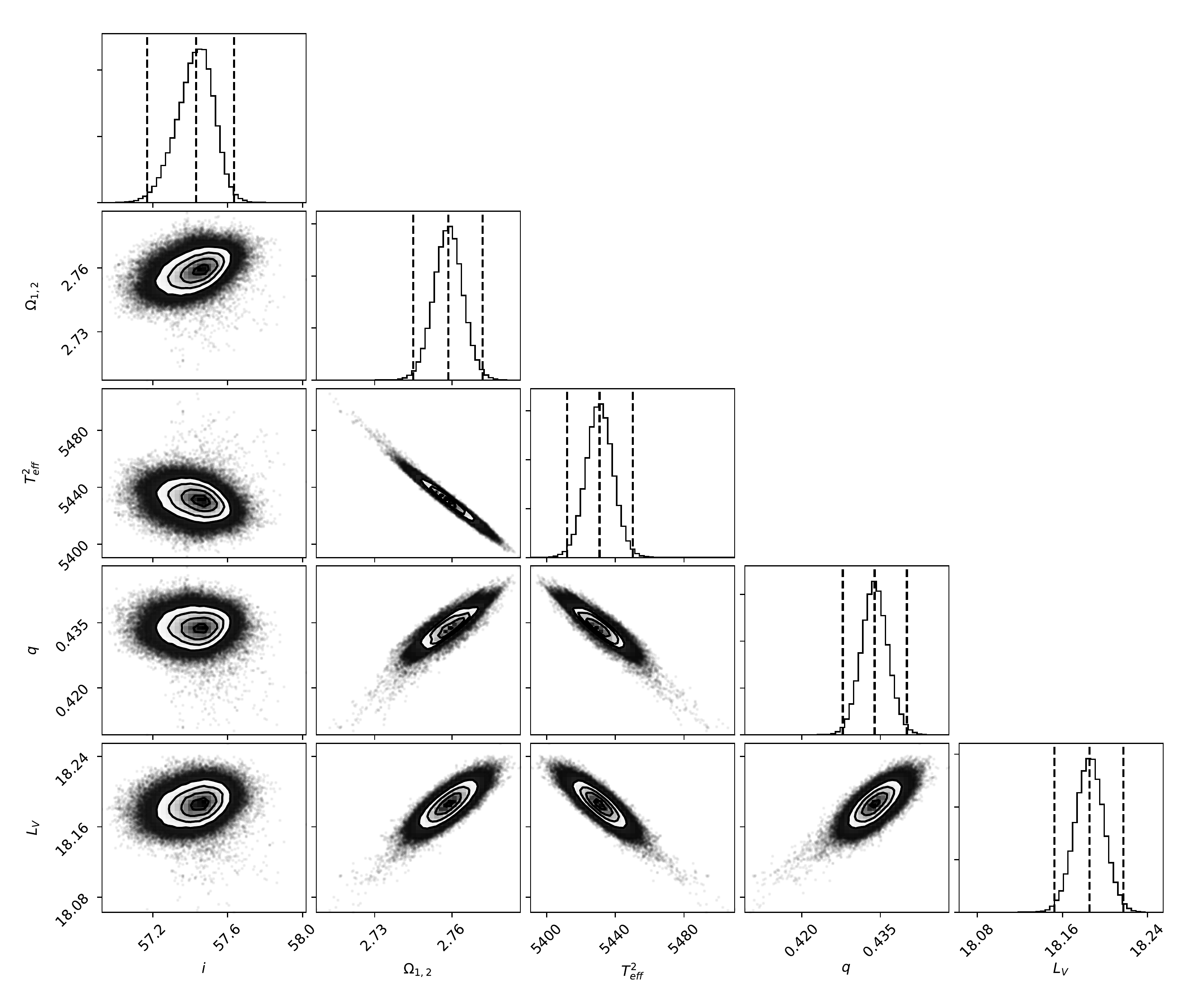}}
\subfigure{\includegraphics[width=8.7cm,height=9.5cm]{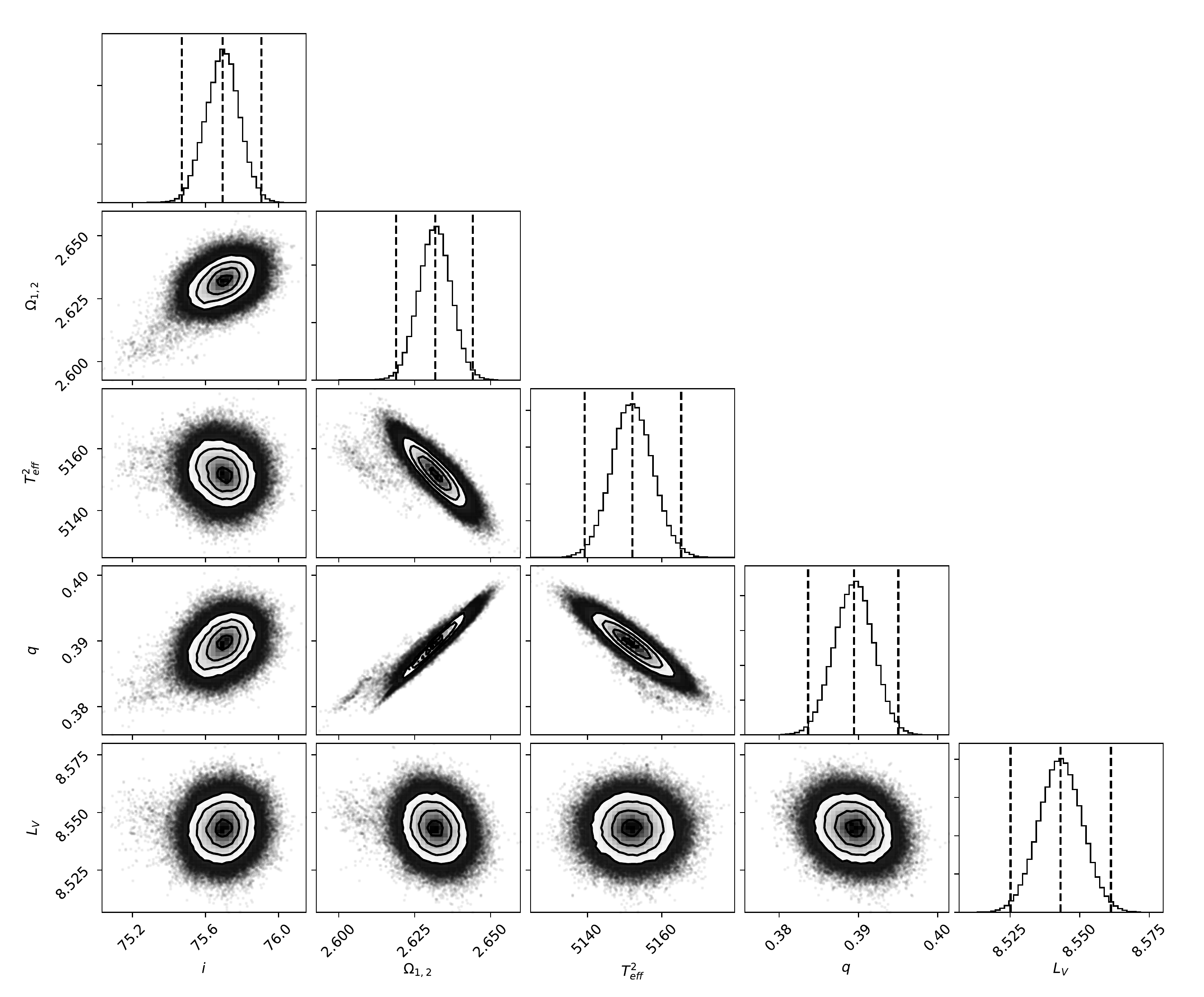}}
\caption{Same as Figure~\ref{corna} but for J1013 (upper left), J1324 (upper right) and J1524 (lower).}
\end{center}
\end{figure*}

\end{appendix}
\end{document}